\documentclass[a4paper,fleqn,usenatbib,useAMS]{mnras}

\usepackage{graphicx}	
\usepackage{amsmath}	
\usepackage{amssymb}	
\usepackage{multicol}        
\usepackage{bm}		
\usepackage{pdflscape}	
\usepackage[utf8]{inputenc}
\usepackage{textcomp}

\newcommand{\kms}{\,km\,s$^{-1}$ } 
\newcommand{\red}{redshift }
\newcommand{\reds}{redshifts }
\newcommand{\los}{line-of-sight }

\usepackage[T1]{fontenc}
\usepackage{ae,aecompl}

\usepackage{txfonts}

\title[The nearby dissociative merging galaxy cluster A3376]{Weak lensing and spectroscopic analysis of the nearby dissociative merging galaxy cluster Abell 3376}

\author[Monteiro-Oliveira et al.] 
  {R.~Monteiro-Oliveira,$^{1,2}$\thanks{Contact e-mail: \href{mailto:rogerionline@gmail.com}{rogerionline@gmail.com}}
  G. B.~Lima Neto,$^1$ 
  E. S.~Cypriano,$^1$ 
  R. E. G.~Machado,$^3$ 
  \newauthor %
  H. V. Capelato,$^{4}$ 
  T. F. Lagan\'a,$^{4}$
  F. Durret$^{5}$ 
  and J. Bagchi$^{6}$
\\
  $^1$Instituto de Astronomia, Geof\'isica e Ci\^encias Atmosf\'ericas, Universidade de S\~ao Paulo, R. do Mat\~ao 1226, 05508-090 S\~ao Paulo, Brazil\\
  $^2$Departamento de Astronomia, Instituto de F\'isica da Universidade Federal do Rio Grande do Sul, Campus do Vale, 91501-970, Porto Alegre, Brazil\\
  $^3$Universidade Tecnol\'ogica Federal do Paran\'a, Rua Sete de Setembro 3165,80230-901 Curitiba, Brazil \\
  $^4$N\'ucleo de Astrof\'isica Te\'orica, Universidade Cruzeiro do Sul, Rua Galv\~ao Bueno, 868, 01506-000 S\~ao Paulo, Brazil\\
  $^5$UPMC-CNRS, UMR 7095, Institut d\textquotesingle Astrophysique de Paris, 75014 Paris, France\\
  $^6$The Inter-University Centre for Astronomy and Astrophysics (IUCAA), Post Bag 4 , Pune 411007, India\\
  }

\date{Accepted 2017 March 28. Received 2017 March 27 ; in original form 2016 December 22}

\pubyear{2017}

\begin{document}
\label{firstpage}
\pagerange{\pageref{firstpage}--\pageref{lastpage}}
\maketitle

\begin{abstract}
The galaxy cluster Abell~3376 is a nearby ($\bar{z}=0.046$) dissociative merging cluster surrounded by two prominent radio relics and showing an X-ray comet-like morphology. The merger system is comprised of the subclusters A3376W \& A3376E. Based on new deep multi-wavelength large-field images and published redshifts, we bring new insights about the history of this merger. Despite the difficulty of applying the weak lensing technique at such low redshift, we successfully recovered the mass distribution in the cluster field. Moreover, with the application of a two-body model, we have addressed the dynamics of these merging system. We have found the individual masses of M$_{200}^{\rm W}=3.0_{-1.7}^{+1.3}\times10^{14}$ M$_{\odot}$ and M$_{200}^{\rm E}=0.9_{-0.8}^{+0.5}\times10^{14}$ M$_{\odot}$. The cometary shaped X-ray distribution shows only one peak spatially coincident with both Eastern  BCG and the A3376E mass peak whereas the gas content of A3376W seems to be stripped out. Our data allowed us to confirm the existence of a third subcluster located at the North, $1147\pm62$ kpc apart from the neighbour subcluster A3376E and having a mass  M$_{200}^{\rm N}=1.4_{-1.0}^{+0.7}\times10^{14}$ M$_{\odot}$. From our dynamical analysis, we found the merging is taking place very close to the plane of the sky, with the merger axis just $10^\circ \pm11^\circ$  from it. The application of a two-body analysis code showed that the merging cluster is seen $0.9_{-0.3}^{+0.2}$ Gyr after the pericentric passage and it is currently going to  the point of maximum separation between the subclusters.

\end{abstract}





\section{Introduction}

Mergers of galaxy clusters are the central mechanism in the formation of the large-scale structure of the universe. According to the $\Lambda$CDM model, galaxy clusters live at the top of the halo mass-function, being formed through a continuous sequence of mergers of smaller structures \citep[e.g.][]{merging_book}. These mergers are massive, with subcomponent masses of $\sim$10$^{14}$ M$_\odot$ and occur at high relative velocities ($\sim$2000 km s$^{-1}$) and with small impact parameters. At around $\sim$$10^{64}$ erg, they constitute the highest energy events since the Big Bang \citep{sarazin04}. Many happen in directions along cosmic filaments, in which matter is distributed in the largest scales \citep{merging_book}.

As the only cluster component subjected to plasma physics during the merger, the intra-cluster medium (ICM) presents the most dramatic signatures of these events, in the form of shocks and cold fronts \citep{markevitch_viki07}. At least for a short period of time, hydrodynamical phenomena can displace the sub-clusters' ICM distributions relative to their dark matter halos (DMH). This detachment can be partial \citep[e.g.][]{monteiro-oliveira17} or even total \citep[e.g.][]{clowe04}. Systems that show this feature are known as dissociative mergers \citep{dawson}.

These shocks induced on the ICM have, in general, Mach numbers $\mathcal{M}\lesssim 3$ \citep[e.g.][]{akamatsu12} and can (re)accelerate relativistic electrons that will emit diffuse, non-thermal synchrotron radiation in radio. These emissions can be then detected but they will not be associated to a single source \citep[e.g.][]{merging_book}. They are the so-called radio relics, which are related to disturbed systems \citep[e.g.][]{cassano10}, although they are not always present in merging clusters.

\begin{figure*}
 \begin{center}
\includegraphics[width=\textwidth, angle=0]{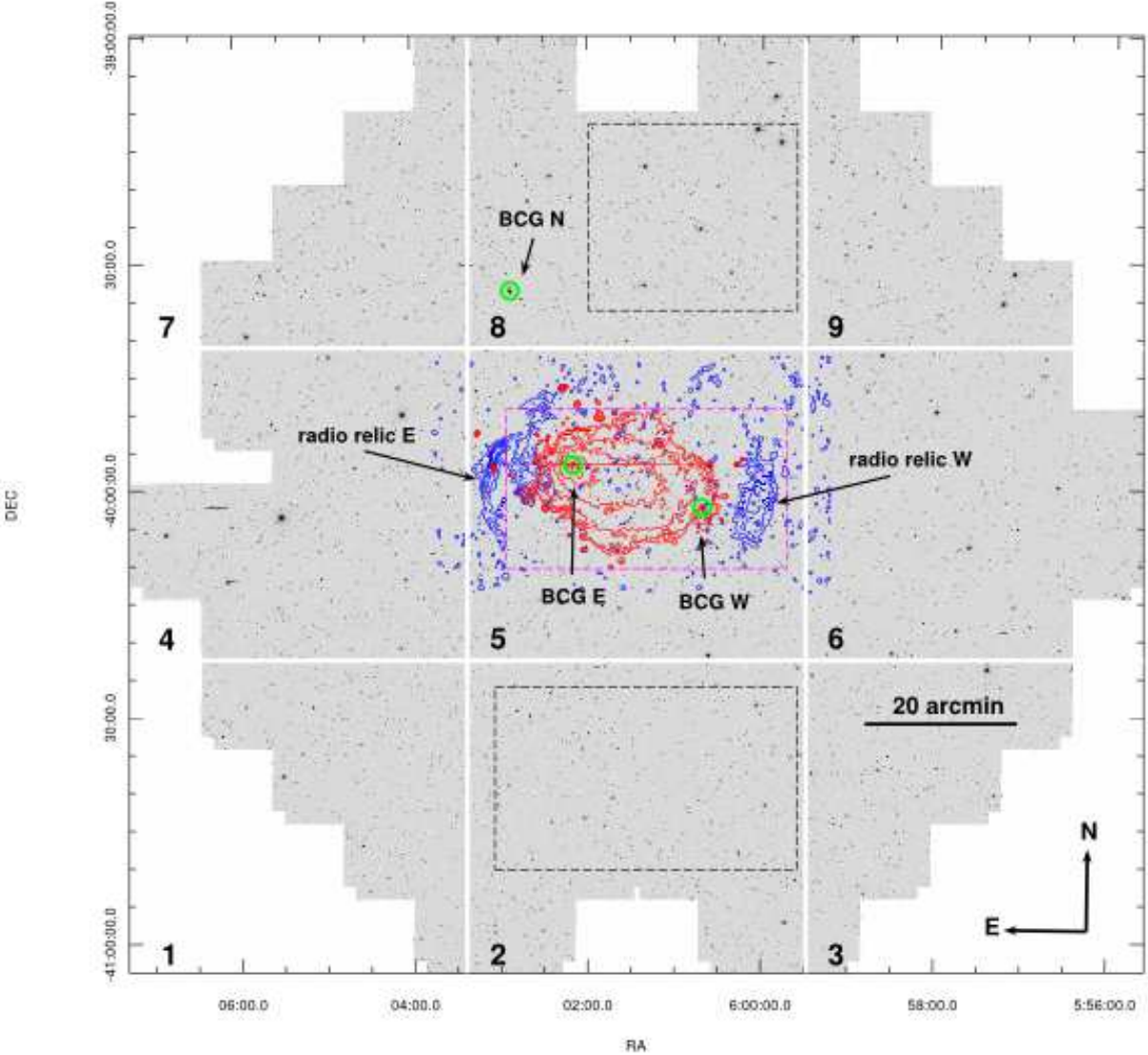}
\caption{Optical $r'$ image of A3376 obtained with the Blanco/DECAM imager. Overlaid are the X-ray emission contours as seen by the satellite XMM-\textit{Newton} ({\it red}) and the radio emission mapped by the VLA telescopes  ({\it blue}). The BCGs, including the Northern candidate are marked by {\it green} circles. The cluster X-ray emission has a morphology similar to a comet whose main axis is nearly aligned with the segment along the BCGs W \& E and the tail points toward the brightest one. In the outskirts, we have two diffuse radio emissions not associated with any single sources: the radio relics. As the image is large ($\sim$2.2 deg$^2$), it was divided into nine frames aiming for the weak lensing analysis. The {\it dashed} boxes show the regions considered for the statistical subtraction to identify the cluster red sequence galaxies: whereas these galaxies are supposed to prevail in the central {\it magenta} region, it is expected that their contribution diminish inside the {\it black} regions. }
\label{fig:A3376.field}
\end{center}
\end{figure*}

The galaxy cluster Abell 3376 (hereafter A3376) is a nearby \citep[$\bar{z}=0.046$;][]{strublerood99} merging system, as revealed by its disturbed ICM \citep{ebeling96,flin06} and the presence of two radio relics at the cluster outskirts \citep{bagchi06,kale12, george15}. The X-ray surface brightness is very high, corresponding to an X-ray luminosity at the cluster redshift of $L_{X[0.5-2.0\ \rm keV]} = 5.72\times 10^{43}$ erg s$^{-1}$ \citep[][]{parekh15} having a comet-like morphology, with its major axis closely aligned with the line connecting the two BCGs, as shown in Fig.~\ref{fig:A3376.field}. For this system, \cite{akamatsu12}  obtained $\mathcal{M}=2.91\pm0.91$ leading to a shock velocity upper limit $v_{\rm shock}<2000$ km s$^{-1}$, believed to have been generated 0.32 Gyr ago. These radio relics are separated by $\sim$2 Mpc and are nearly perpendicular to the line along the BCGs. Furthermore, \cite{george15} argue that they are consistent with a shock propagation initiating $\sim$~0.37 Gyr ago.

For its optical counterpart, \cite{ramella07} have found in the WIde-field Nearby Galaxy-cluster Survey \citep[WINGS;][]{fasano06} two overdensities in the galaxy projected positions that match the BCGs location. They have also reported mild evidence for a third galaxy concentration, but since it was very near the image border, more conclusions could not be drawn. In fact, there is a bright red galaxy located in that region, as revealed by our large field-of-view image. Moreover, \cite{durret13} have shown that the galaxy B-luminosity function is notoriously disturbed, as expected for post merger systems. Examining \red catalogues available in the literature, they found no clear bimodal scenario in the BCGs region as pointed by \cite{ramella07}. 

Using $N$-body hydrodynamical simulations, \cite{machado13} had proposed a scenario of a post minor-merger (mass ratio around 1:6--1:8) whose pericentric passage would have happened 0.5 Gyr before the observed configuration. Due to the unavailability of lensing data at the time, those authors had considered the BCGs ($\sim$970 kpc apart from each other) as tracers of the dark matter halo centroids. Also due to the lack of lensing analysis, the total cluster mass was only known from the galaxies velocity dispersion \citep[$M_{\rm vir}\approx5\times10^{14}$ M$_\odot$;][]{girardi98}. The velocity dispersion, however, is known to be boosted in epochs close to pericentric passages \citep[e.g.][]{pinkney,monteiro-oliveira17}, which can lead to mass overestimation.


In this context, this work presents the first weak lensing analysis of A3376. From deep $r'$ images taken from the  Dark Energy Camera (DECAM) mounted on the Blanco Telescope, we derived weak lensing masses. Using a \red catalogue from literature, we have assigned member galaxies to their respective subclusters and performed their dynamical analysis. Subsequently, with both mass and dynamical information combined, we have addressed the merger scenario of A3376 assuming a two-body dynamical model \citep{dawson}. Finally, photometric and spectroscopic data allowed us to confirm the existence of a third subcluster, as previously suggested by \cite{ramella07}, north of the already known merging system (see Fig.~\ref{fig:A3376.field}).

Although weak lensing has been thoroughly used to probe galaxy clusters, its application to low redshift systems brings an additional challenge. The lensing signal is closely related to the (angular diameter) distance between the lens and background source galaxies, penalising clusters at very low \reds \citep[e.g.][]{spinelli12}.

This paper is organised as follows. In Section~\ref{sec:weak.lensing} we present the weak lensing analysis, from the description of the data up to the mass measurements.  The dynamical overview based on the galaxy \red analysis, can be found in Section~\ref{sec:dynamical.analysis}. The proposed merger scenario for A3376 is described in Section~\ref{sec:merger.scenario}. All results obtained are discussed in Section \ref{sec:discussion} and summarised in Section \ref{sec:summary}.

Throughout this paper we adopt the following cosmology: $\Omega_m=0.27$, $\Omega_\Lambda=0.73$, $\Omega_k=0$, and $H_0=70$~km~s$^{-1}$ Mpc$^{-1}$.  At the mean cluster redshift of $z=0.0461$ we then have $1$ arcsec equals $0.905$~kpc, the age of the Universe $13.2$~Gyr, and an angular diameter distance of $186.4$~Mpc.

\section{Weak lensing analysis}
\label{sec:weak.lensing}

\subsection{A very quick review}
\label{sec:wl.basic}

Weak gravitational lensing is a powerful technique to recover mass distributions of galaxy clusters that works by measuring the distortion caused on the background galaxy shapes. Basically, the lens phenomenon in the weak regime\footnote{The weak regime corresponds to $\kappa\ll 1$.} can be formalised as an isotropic convergence effect $\kappa$ plus an anisotropic distortion described by the shear $\gamma=\gamma_1+i\gamma_2$ where
\begin{equation}
\gamma_1 = |\gamma| \cos(2\theta)\, 
\quad\mbox{,}\quad
\gamma_2 = |\gamma| \sin(2\theta) \,  
\label{eq:gamma.def} 
\end{equation}
and $\theta$ is the shear direction. Mathematically, both convergence and shear quantities are related to the second derivatives of the 2D-projected gravitational potential \citep[e.g.][]{mellier99,schneider05}.

The convergence corresponds to the projected surface mass density of the lens,
\begin{equation}
\kappa = \frac{\Sigma}{\Sigma_{cr}}\,,
\label{eq:kappa}   
\end{equation}
where we define the lensing critical density as
\begin{equation}
\Sigma_{cr} = \frac{c^2}{4\pi G}\frac{D_s}{D_{ds} D_d}\,,
\label{eq:Sigmacr}   
\end{equation}
with $D_s$, $D_{ds}$ and $D_d$ being the angular diameter distances to the source, between the lens and the source, and to the lens respectively. The signal strength depends, then, on the distance between the lens and background objects. Together with the known distribution of galaxies that can be background sources for clusters \citep[$\bar{z}_{\rm back}\approx1$ per deep surveys; e.g.][]{cfhtls}, it is expected that galaxy clusters located at intermediate \reds ($z\sim0.25$--$0.50$) are the most efficient deflectors \citep[e.g.][]{spinelli12}, whereas nearby lenses (as A3376)  will display a very low signal.


In the absence of a massive deflector in the foreground, a sample of distant galaxies will exhibit null average ellipticity, $\langle e \rangle\approx0$. The effect of the lens is to induce on this sample a coherent ellipticity, that corresponds to an effective shear $g$,

\begin{equation}
\langle e \rangle \simeq g \equiv \frac{\gamma}{1-\kappa}\mbox{,}
\end{equation}
represented by a spin-2 tensor (as well as $\gamma$ and $e$). Normally, these quantities are defined in terms of its cross component following Cartesian x--y directions, e.g. $g_+$, and another one $45^\circ$ in relation to that, $g_\times$.

\subsection{Imaging data}
\label{sec:imaging}

Our imaging data were taken with the Dark Energy Camera (DECAM) of the Victor Blanco 4m-telescope (Proposal ID: 2013B-0627; PI: Gast\~ao Lima Neto) within the SOAR Telescope time exchange program. Its large FoV ($\sim$2 deg$^2$) makes it suitable for lensing analysis of a close system, which is expected to cover a large area on the sky. Observational details are presented in Tab.~\ref{tab:imaging}. Before being delivered to the PI, the images were reduced and astrometrically calibrated by the telescope staff \citep{valdes14}.

\begin{table}
\begin{center}
\caption{Imaging data characteristics}
\begin{tabular}{lcc}
\hline
\hline
Band & Total exposure (h) & Seeing (arcsec) \\
\hline
$g'$  & 0.88 &  1.20 \\
$r'$  & 1.66 &  1.20  \\
$i'$  & 0.55 &  1.16 \\
\hline
\hline
\end{tabular}
\label{tab:imaging}
\end{center}
\end{table}

Unlike other telescopes, Blanco allows only observations in classical mode. Unfortunately, our science observations were made under non-photometric conditions, making it difficult to calibrate in the AB system from standard star fields \citep{adelman-mccarthy} observed in the same night. We overcame this by using tabulated first-order extinction coefficients provided by the telescope staff (Alistair Walker in private communication).

Photometric catalogues were created with {\sc SExtractor} \citep{sextractor} using the double mode detection based on the deepest $r'$ band. Galaxies were selected with two complementary criteria: bright objects were considered galaxies if they had their CLASS\_STAR index smaller than 0.8, whereas the faintest, $r'\geq 19$, galaxies had their full width half maximum (FWHM) greater than $1.24$ arcsec.

\subsection{Finding the cluster member galaxies}
\label{sec:red.sequence}

As pointed by \cite{dressler80}, red elliptical galaxies are dominant in the inner region of galaxy clusters, acting as good tracers for these structures. Although cluster member galaxies are not used in weak lensing analysis, their identification is vital in order to exclude them from the background galaxy catalogue, avoiding the addition of noise to the lens model.

The red cluster members have homogeneous photometric properties, occupying a  well defined {\it locus} in the colour-colour (CC) diagram \citep[e.g.][]{med10}. We have used the $r'-i'$ versus $g'-r'$ diagram then to determine this locus through statistical subtraction. To this purpose, we have considered two different regions: a central one ({\it magenta} dashed box in Fig.~\ref{fig:A3376.field}), where we expect most cluster members to be, and two distant regions  ({\it black} dashed boxes) where the cluster galaxy counts are negligible compared to the field. We found that the galaxy counts in the central region are higher than those in the distant regions up to $r'=19.5$. We will therefore take this limit for the cluster member detection.

\begin{figure}
\begin{center}
\includegraphics[width=\columnwidth, angle=0]{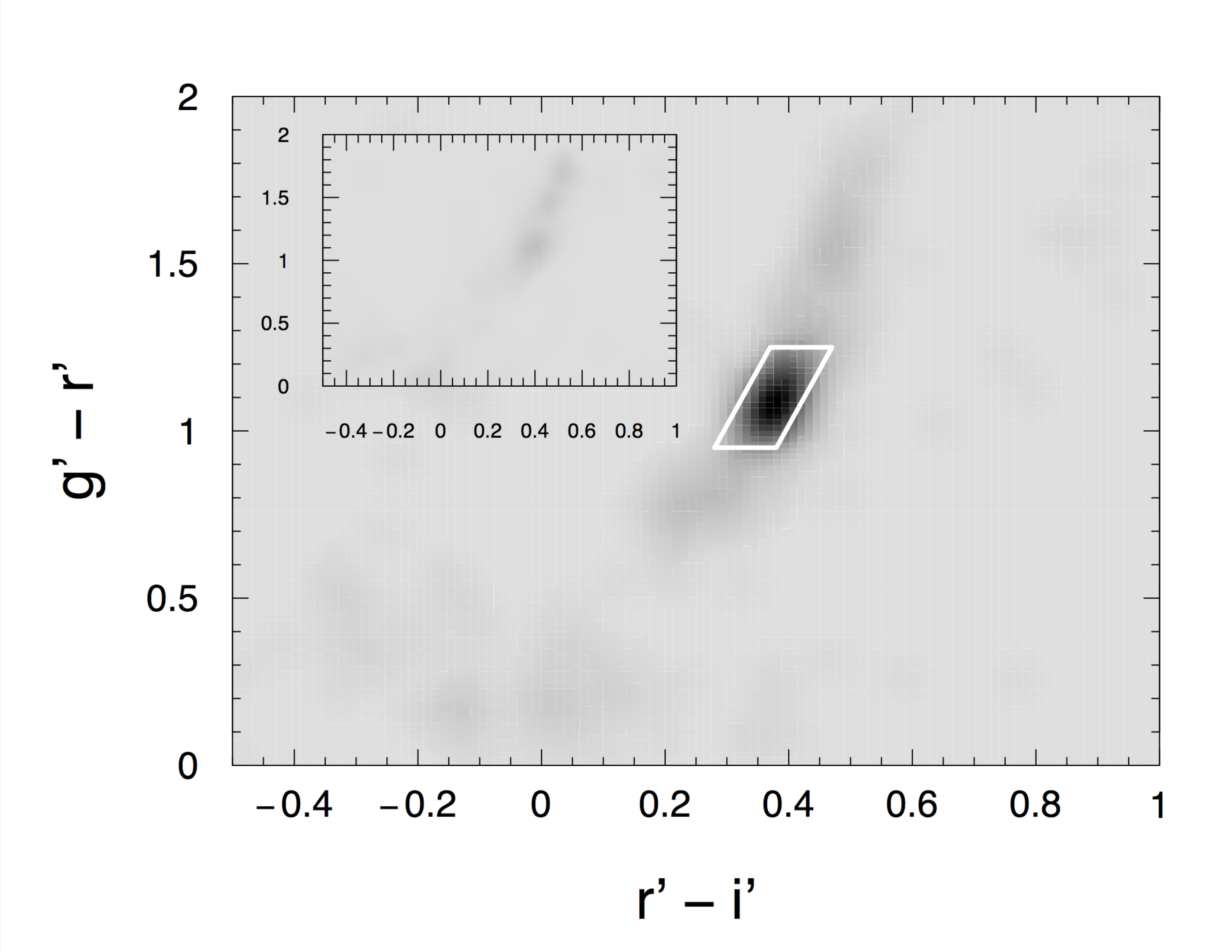}
\caption{Colour-colour diagram  of the galaxies located inside the central ({\it magenta}) box of the Fig.~\ref{fig:A3376.field} and with $r'\leq 19.5$. The evident density peak is related to the red cluster member galaxies. The inset panel shows the same plot for the galaxies located inside the two distant ({\it black}) boxes of the same figure. In both plots the linear grey scale are the same. The {\it white} polygon corresponds to the red cluster member {\it locus}.}
\label{fig:color.color}
\end{center}
\end{figure}
\begin{figure}
\begin{center}
\includegraphics[width=\columnwidth, angle=0]{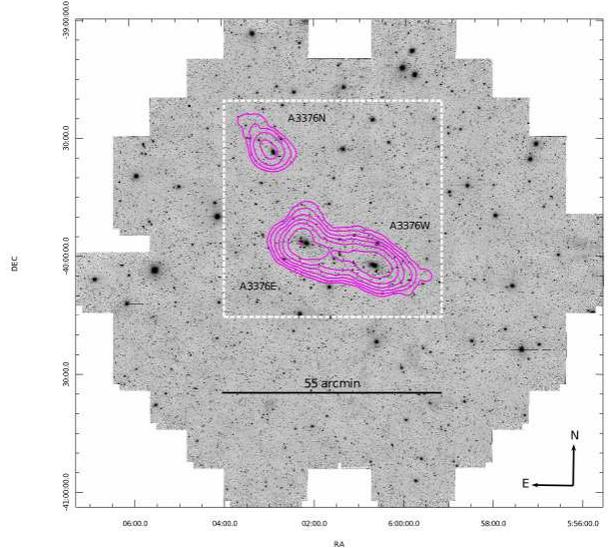}
\caption{The magenta contours correspond to the numerical density distribution weighted by the $r'$ luminosity for the 294 red sequence galaxies selected through colour-colour statistical subtraction. The ``new'' structure A3376N is centred on a bright red luminous galaxy (hereafter BCG N) $\sim$25 arcmin ($\sim$1.4 Mpc) apart from the nearest BCG E. The {\it white} dashed box delimits the area where we will carry further studies on the mass distribution.}
\label{fig:luminosity}
\end{center}
\end{figure}

In Fig.~\ref{fig:color.color} we show the CC diagrams of central and peripheral regions (inset) of the image (Fig.~\ref{fig:A3376.field}). The former presents a clear excess over the latter that is related to the cluster members. By subtracting the number densities in a grid over the colour-colour space, this excess becomes even more evident and thus we define the region marked in the diagram around it.  After this process, we have selected 294 members within $r'\leq19.5$.  Their numerical density distribution weighted by the $r'$-luminosity can be seen in the Fig.~\ref{fig:luminosity}.

The numerical density distribution weighted by the  $r'$-luminosity  shows a tri-modal  behaviour being the most pronounced feature related to the merging subclusters A3376W \& A3376E. At the North, as was previously suggested by \cite{ramella07}, we have found a sample of galaxies whose colours are compatible with the main A3376 system. The final confirmation about its pertinence or not to A3376 will come from the analysis of the galaxy \reds in this region. (Sec.~\ref{sec:dynamical.analysis}).

\subsection{Shape measurements}
\label{sec:shape.measu}

Finding precise corrections for the point spread function (PSF) is a crucial task for the success of shape measurements of faint objects. Specially so in large cameras as DECAM, the PSF exhibits large variation across the field, which has to be taken into account. We mapped the PSF independently on each $44\times41$ arcmin frame (\#1:9 in Fig.~\ref{fig:A3376.field})  by selecting bright, unsaturated stars. Being point-like objects, all deviation from circularity detected in their shapes are due only to instrumental/atmospheric distortion of the PSF. For DECAM, the PSF is also magnitude dependent \citep[``the brighter, the fatter''; e.g][]{melchior15} reflecting the tenuous flux variance on the efficiency of the registration charges in the CCD \citep{antilogus14}.

Shape measurements of PSF distortion across the field were made with a Bayesian publicly available code {\sc im2shape} \citep{im2shape}\footnote{\texttt{http://www.sarahbridle.net/im2shape/}}. This code works by modelling galaxies as a sum of Gaussians with an elliptical basis. On the other hand, stars are modelled as a single Gaussian profile and no deconvolution is performed  to keep the PSF parameters, which are the ellipticity components ($e_1$, $e_2$), and the FWHM. We have used the {\sc R} \citep{R} function thin plate spline regression \citep[{\sc Tps};][]{fields} to spatially interpolate the discrete set of PSF parameters allowing the model to consider also the corresponding magnitudes. For the sake of quality, we iterated this process three times, each time removing the 10\% of objects with the largest absolute residuals\footnote{We have adjusted $df=200$ parameters, leading to smooth PSF parameter surfaces.}.

\begin{figure*}
 \begin{center}
\includegraphics[width=0.8\textwidth, angle=0]{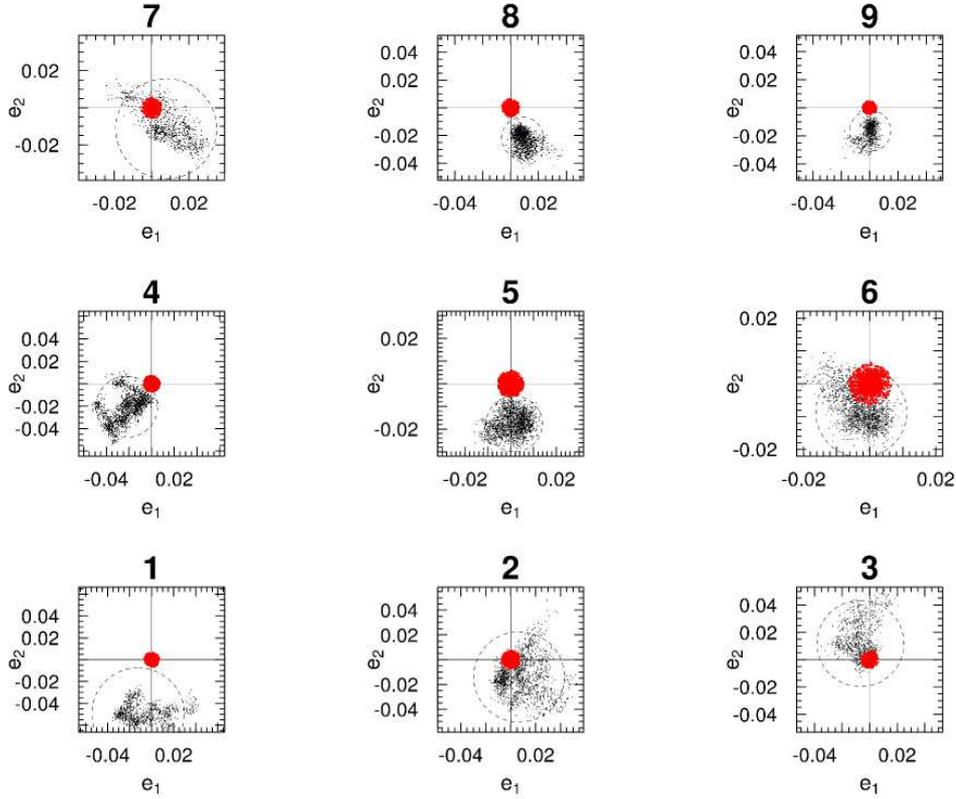}
\caption{Raw distribution of the ellipticity in Cartesian components $e_1$ and $e_2$ measured on selected unsaturated bright stars by {\sc im2shape} ({\it black} points). Aiming at an accurate PSF correction, we have performed our analysis individually for each frame as defined in Fig.~\ref{fig:A3376.field}. After the iterative process to quantify the PSF effect was finished, we computed the residuals between the model and the data. As shown by the {\it red} points, they have good concordance with zero regardless of the initial data spread. The {\it dashed} circles comprise 95\% of the data. }
\label{fig:psf.correction}
\end{center}
\end{figure*}

The final measured ellipticities and the corresponding residuals after the spatial and flux interpolation are shown in Fig.~\ref{fig:psf.correction}. In the end, all galaxies were modelled as a sum of two Gaussians with an identical basis and the local was PSF deconvolved. Furthermore, we removed galaxies with large ellipticity errors ($\sigma_e>2$) and with evidence of contamination by nearby objects.

\subsection{Source selection}
\label{sec:source.selection}

The shape deformation on background galaxies is used to infer the mass distribution which acted as a gravitational lens. In this sense, it is strictly necessary to perform a careful selection of background galaxies in order to diminish the contamination by both foreground and cluster member galaxies. Because of its proximity ($\bar{z}=0.046$), it is expected that the galaxy sample in the A3376 field is strongly dominated by background ones. For a comparison, the Canada-France-Hawaii Telescope Legacy Survey (CFHTLS)\footnote{\texttt{http://www.cfht.hawaii.edu/Science/CFHTLS/}}, shows that  only 0.14\% of its galaxies have $z\leq0.046$, making the contamination by these galaxies negligible. Thus, we have just excluded the red sequence CC locus and considered as background candidates all galaxies having $r'\geq21$.

Due to the fact that A3376 is located at very low redshift, it is expected that the background structures will appear in the mass map with, at least, comparable or higher signal than the subject galaxy cluster, according to Eq.~\ref{eq:Sigmacr}. In order to overcome this challenging, but unavoidable obstacle, we have concentrated our efforts in extracting the lensing signal of A3376 by reducing the signal coming from high \red structures. Without a photometric redshift catalogue for our images, we tested several cut limits for the fainter galaxies (closely related to high \red ones) aiming to maximise the signal-to-noise near the BCGs location.

For this purpose, we referred to the mass aperture statistic \citep{schneider96} whose S/N is averaged through the measured tangential ellipticity $e_\times$ for the $N_{\theta_0}$ galaxies inside a radius $\theta_0$,

\begin{equation}
{\rm S/N}=\dfrac{\sqrt{2}}{\sigma_e^2}\ \dfrac{\sum_{i=1}^{N_{\theta_0}} e_{\times_i}(\theta_i) Q_{\rm NFW}(\theta_i,\theta_0)}{\left[ \sum_{i=1}^{N_{\theta_0}} Q_{\rm NFW}^2(\theta_i,\theta_0)\right ]^{1/2}} \ \mbox{,}
\label{eq:SN}
\end{equation} 
where $\theta_i$ is the galaxy position and  $\sigma_e$ is the quadratic sum of the measured error and the intrinsic ellipticity uncertainty, estimated as $0.35$ for our data. The NFW filter \citep{schirmer04},
\begin{multline}
\noindent Q_{\rm NFW}(\theta_i,\theta_0)=[1+e^{a-b\chi(\theta_i,\theta_0)}+e^{-c+d\chi(\theta_i,\theta_0)}]^{-1} \times  \\
\dfrac{\tanh [\chi(\theta_i,\theta_0)/\chi_c]}{\pi\theta_0^2[\chi(\theta_i,\theta_0)/\chi_c]}\mbox{,}
\label{eq:nfw.filter}
\end{multline}
where $\chi=\theta_i/\theta_0$, describes approximately the NFW shear profile. \cite{hetterscheidt05} suggest, as optimised parameters for the halo detection maximisation, $a=6$, $b=150$, $c=47$, $d=50$, $\chi_c=0.15$ and $\theta_0=11$ arcmin.

\begin{figure}
\begin{center}
\includegraphics[width=0.8\columnwidth, angle=-90]{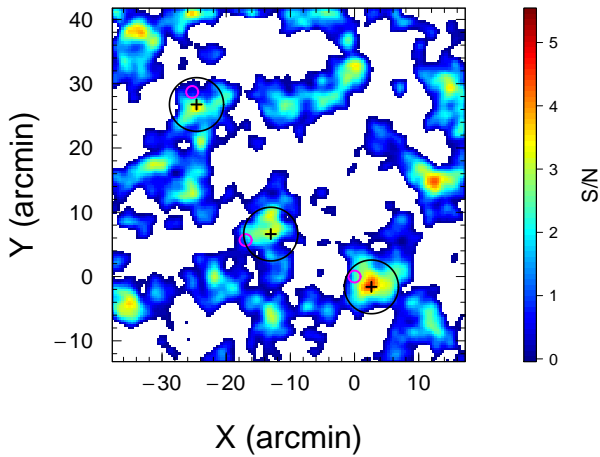}
\caption{Signal-to-noise map obtained through the mass aperture statistic \citep{schneider96} considering a fainter limit of $r_{\rm max}=24.5$. The {\it white} regions correspond to a negative S/N. The BCGs are marked as small {\it magenta} circles whereas their neighbour highest S/N peak (``+'') are marked as large {\it black} circles.  Because of the low redshift, A3376 is surrounded by a lot of background structures.}
\label{fig:sn}
\end{center}
\end{figure}

The S/N map of the central region ({\it white} box in Fig.~\ref{fig:luminosity}) is presented in Fig.~\ref{fig:sn}. As expected, it shows a multi-structure scenario. The highest S/N peaks around the BGCs are reported by the $21\leq r'\leq24.5$ limited sample, hereafter our fiducial background sample. This final background sample had a field density of 12.3 galaxies per arcmin$^{-2}$. The comparison with CFHTLS allowed us to estimate the critical density as  $\Sigma_{cr}=9.6\times10^9$ M$_{\odot}$ kpc$^{-2}$.

\subsection{Mass reconstruction}
\label{sec:mass.reconstruction}

Now, from measured ellipticities, we reconstructed the mass distribution using the code {\sc LensEnt2} \citep{LensEnt2}. Its Bayesian method is based on the maximum entropy algorithm \citep{seitz98}, which works to minimise a $\chi^2$-like statistics based on the comparison between the measured ellipticity field with those the model predicts. The field is smoothed by a Gaussian filter with $\sigma=160$ arcsec which was the  best compromise between the signal-to-noise in the final map. This smoothing is required since each galaxy is not isolated, i.e., they are correlated with those in their neighbourhoods.

\begin{figure*}
 \begin{center}
\includegraphics[width=0.8\textwidth, angle=0]{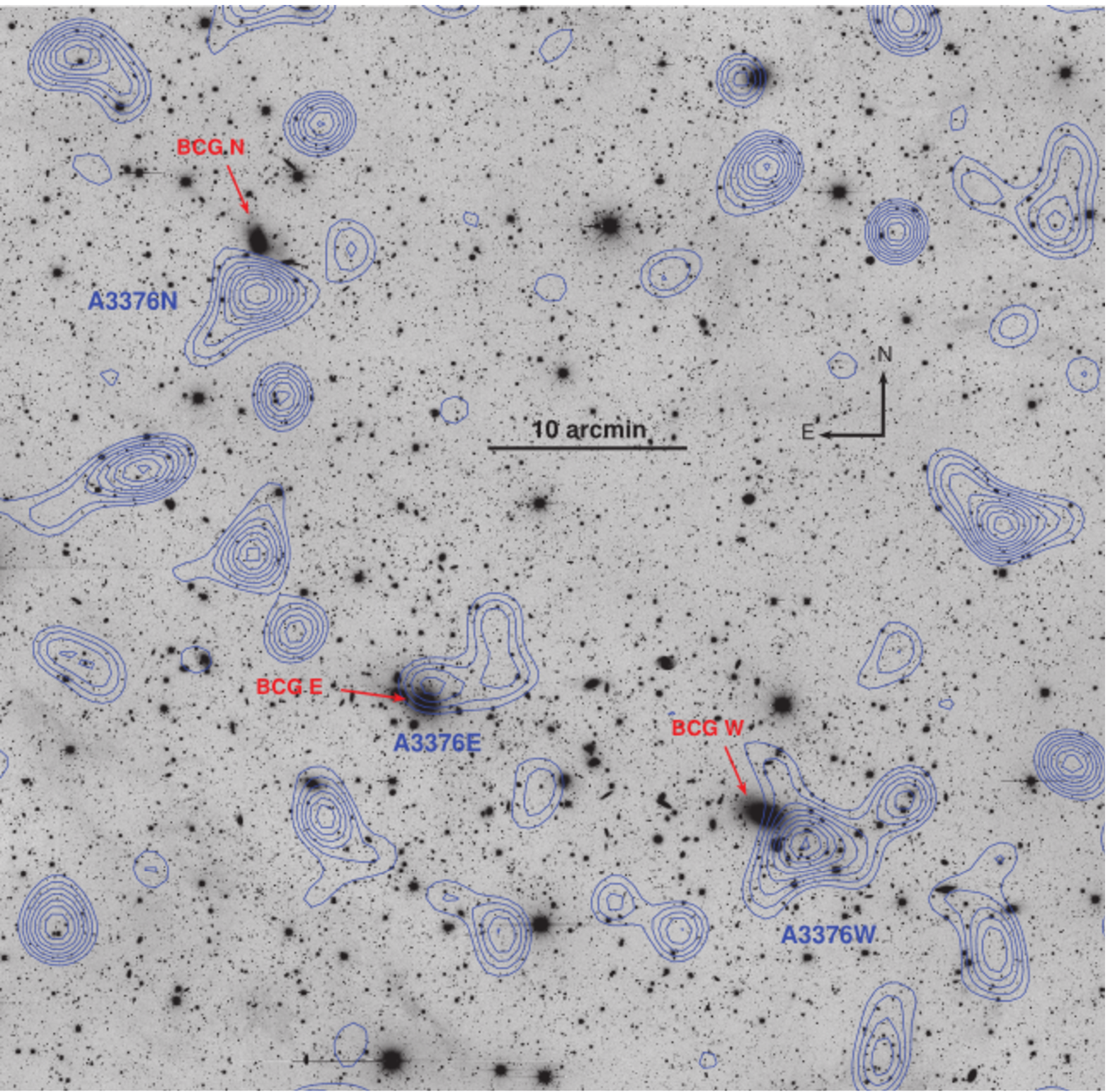}
\caption{Mass distribution ({\it blue} contours) in the central part of A3376 field as obtained by {\sc LensEnt2} overlaid with the $r'$ band image. The contours are linearly spaced by 0.01 within the interval $\kappa \in [0.05:0.15]$. The merging subclusters A3376W \& A3376E were identified as the nearest mass peaks respective to their BCGs.  The same procedure was applied to identify the mass concentration related to the Northern subcluster candidate A3376N.}
\label{fig:mass.contours}
\end{center}
\end{figure*}

The resulting mass distribution is presented in Fig.~\ref{fig:mass.contours}.  The subclusters' mass structures were identified as the closest to their respective BCGs, as suggested by the previous numerical density distribution weighted by the  $r'$-luminosity (Fig.~\ref{fig:luminosity}). As anticipated by the S/N map, the mass distribution shows, beyond the (at least) three A3376 related peaks, a great amount of clumps in the field. Moreover, the A3376E subcluster mass distribution looks blended with another nearby mass concentration. In the following, we apply a method to quantify the mass of separate structures and estimate their central position errors.

\subsection{Mass distribution modelling}
\label{sec:mass.modelling}

The multi-peaked mass distribution brings an additional difficulty to measure the mass of the structures we are interested in. Initial efforts to fit three different NFW profiles for the A3376 related clumps have proven unfruitful because the model did not reach convergence even after many iterations ($\sim$ $1\times10^5$). To properly address this, we had to identify the most relevant structures in the field and then add them into our model.

\begin{figure}
\begin{center}
\includegraphics[width=\columnwidth, angle=0]{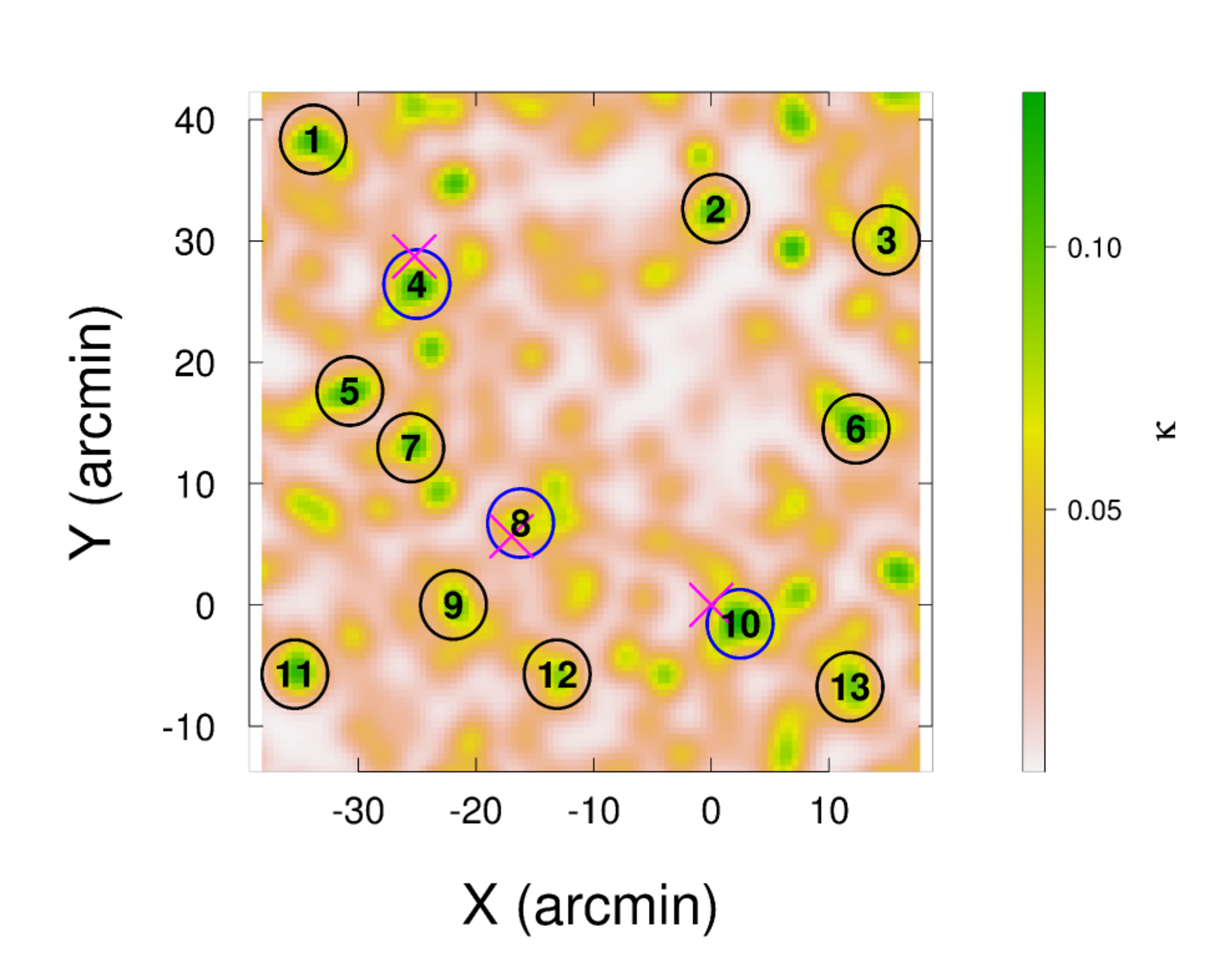} 
\caption{Convergence map showing the most significant mass peaks in A3376 field according to our peak finder method. The circular moving window with $R=2.8$ arcmin found 13 mass peaks above the minimum significance level $1.6\sigma$, as seen in the A3376E mass related clump.The BCGs are marked with the {\it magenta} ``X''. The peaks \#10, \#8 and \#4 correspond, respectively, to the A3376W, A3376E and A3376N subclusters. Our forthcoming model will consider these thirteen peaks simultaneously.}
\label{fig:peak.id}
\end{center}
\end{figure}

We searched for peaks inside a moving circular window 2.8 arcmin in radius. This procedure allowed us to identify two subclusters with significances larger than 3$\sigma$. As a lower limit, we considered the significance of the A3376E mass concentration (1.6$\sigma$). The resulting 13 identified peaks are shown in Fig.~\ref{fig:peak.id}.

Our fiducial model consists of 13 NFW profiles fitted simultaneously, each one with masses as their  unique unknown parameters. In the absence of circular symmetry, we worked with Cartesian components of the effective shear rather than the usual tangential component. The transformation can be done easily just by multiplying $g_\times$ by the lensing convolution kernel.

\begin{equation}
D_1 = \frac{y^2 - x^2}{x^2 + y^2}\, 
\quad\mbox{and}\quad
D_2 = \frac{2xy}{x^2 + y^2} \, , 
\end{equation}
where $x$ and $y$ are the Cartesian coordinates relative to the respective lens centre. 

The halo concentration was fixed based on the \cite{duffy08} mass-concentration relationship,
\begin{equation}
c=5.71\left(\frac{M_{200}}{2\times10^{12}h^{-1}M_{\odot}}\right)^{-0.084}(1+z)^{-0.47},
\label{eq:duffy_rel}
\end{equation}
where the redshift was considered the same for all halos\footnote{We have considered all of them located at the A3376 redshift.}. 

The total effective shear component in each lens position can be written as a contribution of each halo,
\begin{equation}
g_i = \sum_{k=1}^{13}  g_i^{k}\mbox{,}\quad i\in\{1,2\}\mbox{.}
\label{A3376.gt}
\end{equation}
which was compared to the measured ellipticities $e_1$ and $e_2$ through the $\chi^2$,

\begin{equation}
\chi^2=\sum_{j=1}^{N_{{\rm sources}}} \sum_{i=1}^{2}  \frac{(g_i-e_{i,j})^ 2}{\sigma_{int}^2+\sigma_{ obs_{i,j}}^2}\mbox{,}
\label{A3376.eq:chi2}
\end{equation}
where $\sigma_{ obs_{i,j}}$ corresponds to errors reported by {\sc im2shape}, and $\sigma_{int}$ is the uncertainty associated with the sources intrinsic ellipticity distribution, estimated as 0.35. 

In our Bayesian model, the posterior of the problem is given by 
\begin{equation}
{\rm Pr}(\rm M| data) \propto  \mathcal{L}({\rm data}|M)\times\mathcal{P}(M)\mbox{,}
\label{A3376:eq:posterior}
\end{equation} 
where, additionally, we adopted a uniform prior $0<M\leq 8\times 10^{15}$ $M_{\odot}$ to allow the chains to quickly reach the stationary state. Considering the data normally distributed around the model, we can associate the likelihood with the previous $\chi^2$ statistics as
\begin{equation}
\ln \mathcal{L} \propto - \frac{\chi^2}{2}\mbox{.} 
\label{A3376.likelihood.d}
\end{equation}

\subsection{Results}
\label{sec:results}

We evaluated the posterior described in Eq.~\ref{A3376:eq:posterior} by using a MCMC (Monte Carlo Markov chain) algorithm with a simple Metropolis sampler implemented in the {\sc R} package {\sc MCMCmetrop1R} \citep{MCMCpack}. We generate four chains of $1\times10^5$ points allowing an additional $1\times10^4$ points first as ``burn-in'' to ensure the chains fully represent the stationary state. The potential scale factor $R$, as implemented in the {\sc Coda} package \citep{coda}, has shown that the final combined chain is, within 68\% c.l, convergent ($R<1.05$).

Subcluster masses, marginalised over all parameters, are presented in Tab.~\ref{tab:masses}. For the merging system we have obtained M$_{200}^{\rm W}=3.0_{-1.7}^{+1.3}\times10^{14}$ M$_{\odot}$ and M$_{200}^{\rm E}=0.9_{-0.8}^{+0.5}\times10^{14}$ M$_{\odot}$. This corresponds to a total mass M$_{200}^{\rm W+E}=4.1^{+1.5}_{-1.8}\times10^{14}$ M$_\odot$. We have found that the probability of A3376E being the most massive one is of only 9\%. For A3376N, we obtained  M$_{200}^{\rm N}=1.4_{-1.0}^{+0.7}\times10^{14}$ M$_{\odot}$.

\begin{table}
\caption[]{Median and 68\% range of the marginalised masses in sampled in the MCMC chains.}
\label{tab:masses}
\begin{center}
\begin{tabular}{|l|ccc|}
\hline \hline
 Subcluster&    M$_{200}$ ($10^{14}$ M$_{\odot}$)  \\
\hline
A3376W   & $3.0_{-1.7}^{+1.3}$  \\ [5pt] 
A3376E   & $0.9_{-0.8}^{+0.5}$  \\ [5pt] 
A3376N   & $1.4_{-1.0}^{+0.7}$  \\
\hline \hline
\end{tabular}
\end{center}
\end{table}

We can now ask about the relative position of the mass and X-ray distributions. We have used publicly available XMM-\textit{Newton} observations of A3376 made in 2003 (Obs Id. 0151900101; PI  M.~Markevitch) and 2007 (Obs Id. 0504140101; P.I. M.~Johnston-Hollitt). We have previously used these observations in \cite{machado13} in order to constrain the dynamical modelling of A3376.

For completeness, we summarise here the procedure we have used to make the X-ray intensity map with the EPIC camera data. The XMM-\textit{Newton} data were processed through the SAS\footnote{Science Analysis System 11, \texttt{http://xmm.esac.esa.int/}} pipeline, following the procedure recommended in their site, and keeping only events in the field of view with PATTERN $\le$ 12 (MOS 1 and 2) and PATTERN $\le 4$ (pn). Finally, we have used light curves in order to exclude high particle background time intervals from the observations. With the cleaned event files, we have made exposure map corrected images in broad band (0.5--8.0 keV) and then combined all images (from both observations and for all EPIC cameras) into a single intensity map mosaic.

\begin{figure*}
 \begin{center}
 \includegraphics[angle=0, width=1.0\textwidth]{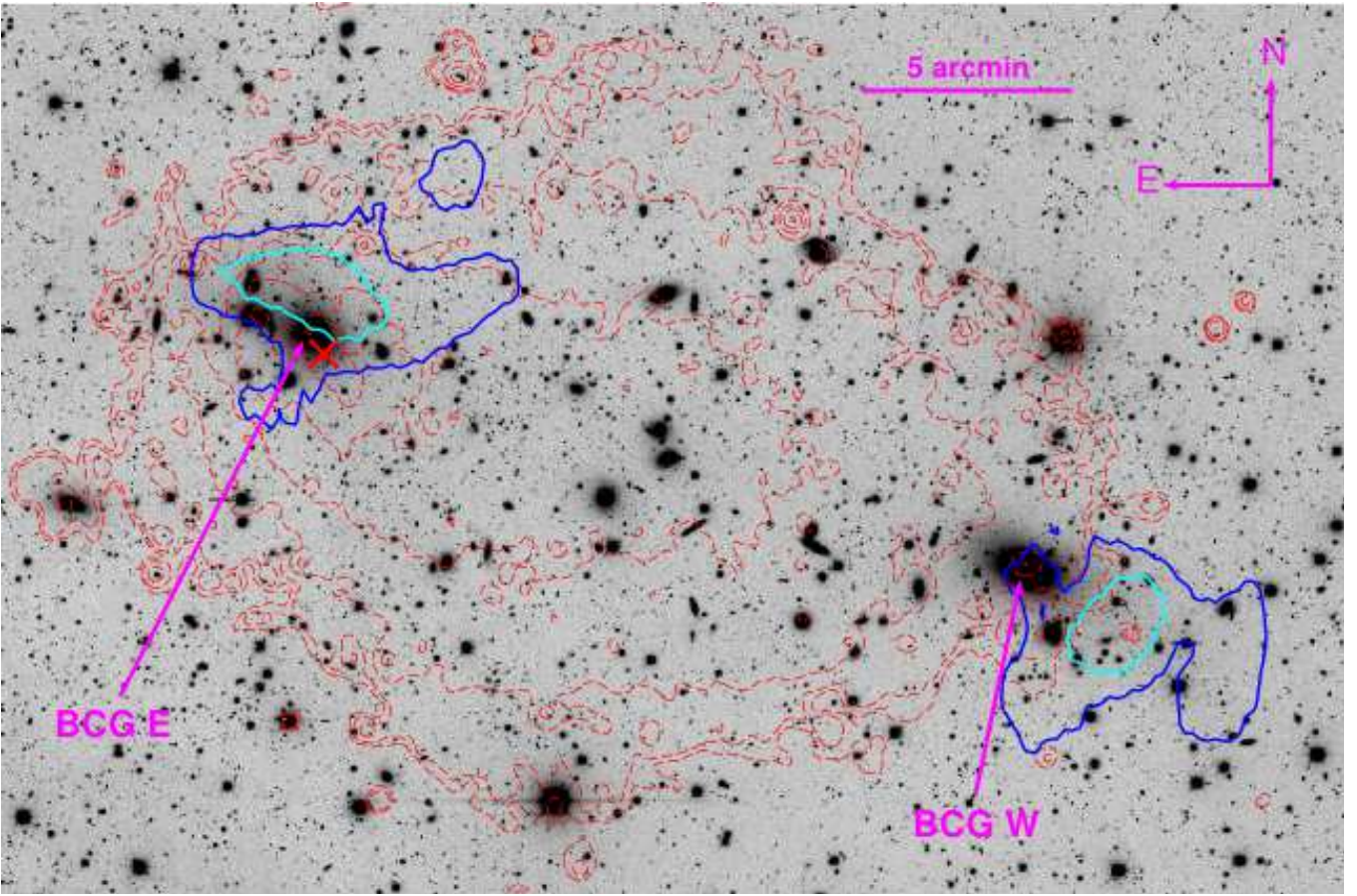} 
 \caption{Optical $r'$ band overlaid with XMM-{\it Newton} X-ray contours ({\it red} dashed lines) and the confidence contours of 68\% c.l. ({\it cyan}) and 95\% c.l. ({\it blue}) of the mass centre positions. The X-ray distribution is unimodal (the {\it red} ``X'' marks it peak) with the peak related to the A3376E subcluster. Both BCG and X-ray position are comparable within 95\% c.l. with the mass centre. The same BCG behaviour in relation to the mass centre is reported on A3376W.}
 \label{fig:A3376:mass.rx}
 \end{center}
 \end{figure*}

The multi-peaked field prevents the simultaneous centre position and mass modelling \citep[e.g.][]{monteiro-oliveira17}. To estimate the uncertainties on the mass centre positions, we performed $1\times10^4$ bootstrap re-samplings of the ellipticity field. For each one, we have reconstructed the mass distribution (as described in Sec.~\ref{sec:mass.reconstruction}) and mapped the position of the nearest peak in relation to those previously identified (Fig.~\ref{fig:peak.id}). The confidence contours are presented in Fig.~\ref{fig:A3376:mass.rx}.

The comet-like X-ray morphology reveals just one more concentrated location close to the BCG E. Their positions are, within 95\% c.l., in agreement to both BCG and mass centre location. On the other hand, the subcluster A3376W seems to have had its gas stripped-off during the collision. According to the simulations of \cite{machado13}, this feature could be understood as the result of a collision between two haloes having different central gas concentrations. In that scenario, the smaller subcluster A3376E is the more concentrated one. It is able to retain its gaseous core after the collision.

\section{Dynamical analysis}
\label{sec:dynamical.analysis}

\subsection{Spectroscopic data}
\label{sec:spectroscopic}

\begin{figure}
\begin{center}
\includegraphics[width=\columnwidth, angle=0]{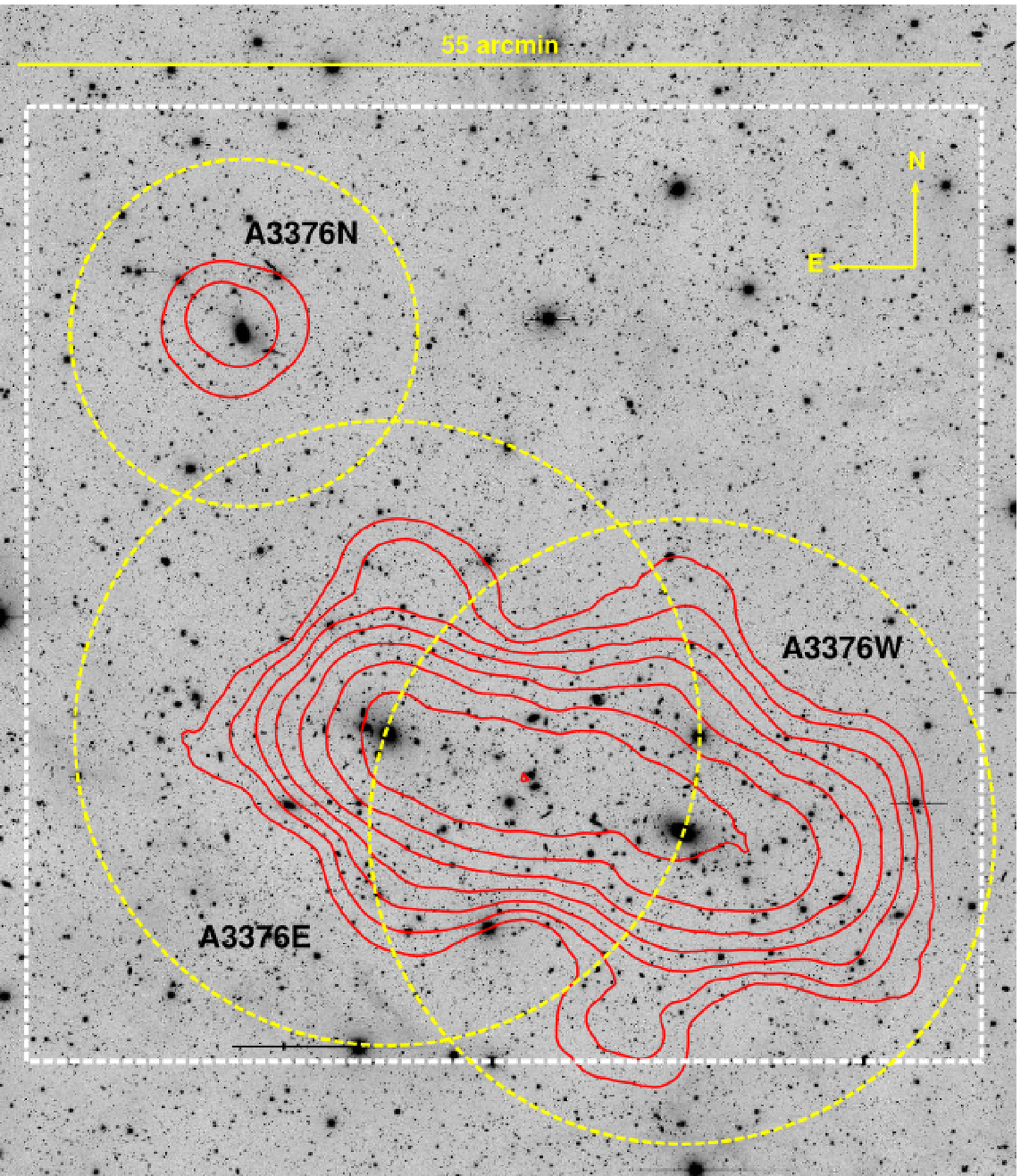}
\caption{Density logarithm contours ({\it red}) of the 173 spectroscopic galaxy members found by the 3$\sigma$-clipping procedure \citep{3sigmaclip}. These galaxies shows good match with the already selected photometric red sequence galaxies. In our analysis, we have considered only the densest regions of the field, highlighted by the {\it yellow} circles centred on the respective BCG. For A3376N we selected galaxies inside a $R\leq10$ arcmin circle, whereas for A3376W\& A3376E the circles had $R=18$ arcmin. Despite the intersection between A3376N and A3376E, there is no galaxy in this region. The other 31 members galaxies are spread along the field.}
\label{fig:density.ned}
\end{center}
\end{figure}

\begin{figure}
\begin{center}
 \includegraphics[width=0.7\columnwidth, angle=-90]{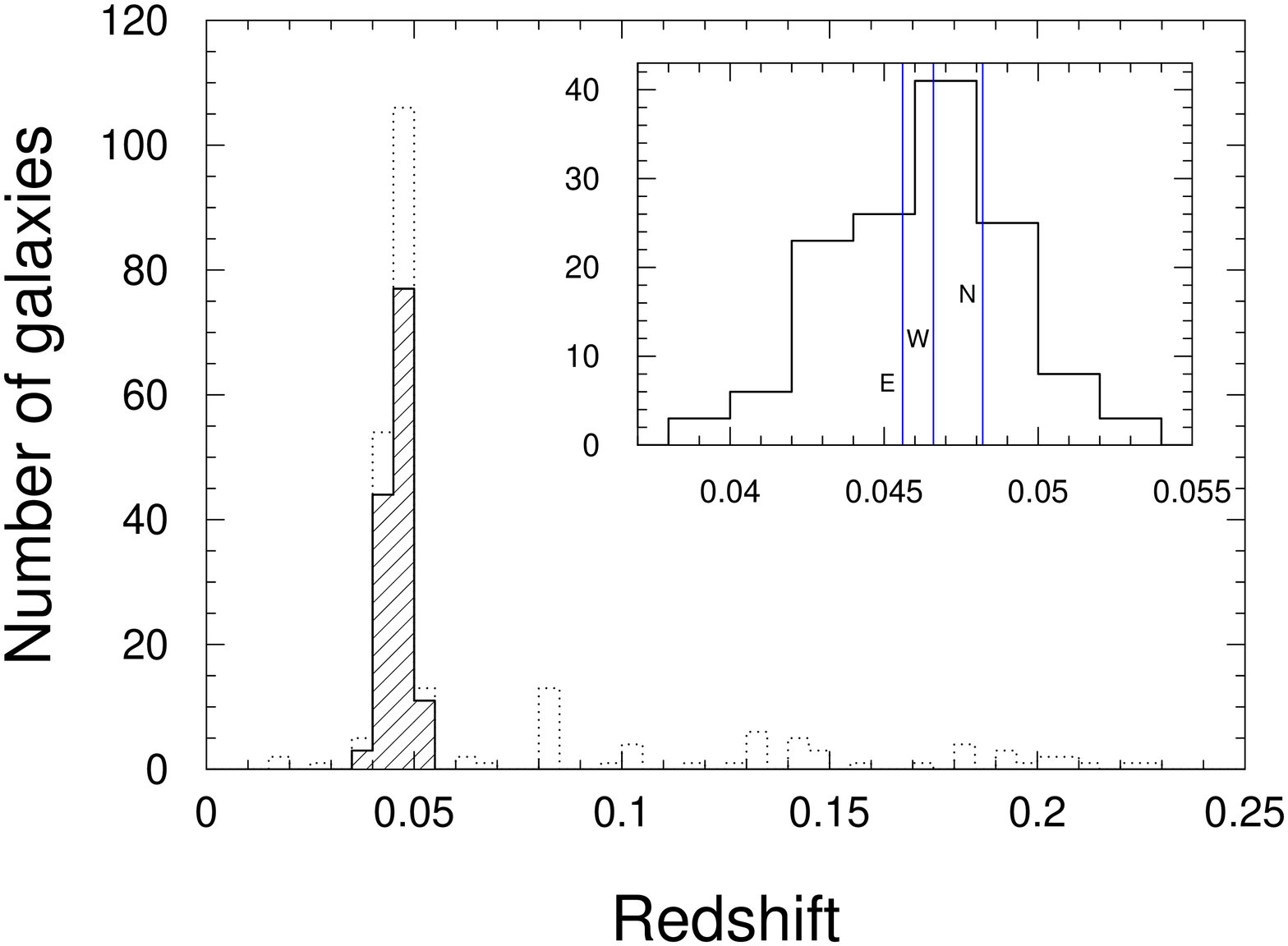}
 \caption{Redshift distribution of the 239 galaxies in A3376 field ({\it dotted} lines). The inset panel, corresponding to the hatched histogram, shows  the 132 members of the merging system A3376W \& A3376E. This sample comprises  the galaxies located inside the combined circular regions with $R\leq18$ arcmin centred on the BCGs W \& E selected after the application of the 3$\sigma$-clipping. This sample has $\bar{z}=0.0461\pm0.0003$ and $\sigma_v/(1+\bar{z})=835\pm45$ km s$^{-1}$. The  vertical lines are showing each BCG redshift. Four galaxies having $z>0.25$ were omitted for clarity.}
 \label{fig:redshift.distribution}
\end{center} 
\end{figure}

A subsequent search on \textit{NASA Extragalactic Database} (NED) revealed 239 galaxies with available redshifts in the A3376 DECAM field. From these, 173 were spectroscopic members after implementing a 3$\sigma$-clipping procedure \citep{3sigmaclip}. Their spatial distribution, seen in Fig.~\ref{fig:density.ned}, follow closely the photometric red sequence distribution (Fig.~\ref{fig:luminosity}). Their respective redshift distribution is shown in Fig.~\ref{fig:redshift.distribution}. For the analysis, we considered only the densest regions, selecting the galaxies located inside the {\it yellow} circles in Fig.~\ref{fig:density.ned} as the fiducial sample.

\subsection{A3376N}
\label{sec:dynamical.analysis.N}

The 3$\sigma$-clipping applied over the whole galaxy sample has shown that the galaxies associated to the subcluster candidate have \reds comparable with those found in the main system. The 7 identified galaxies have $\bar{z}=0.0472\pm0.0005$ with a velocity dispersion of $\sigma_v/(1+\bar{z})=414\pm82$ km s$^{-1}$. 

\subsection{A3376W \& A3376E}
\label{sec:dynamical.analysis.WE}

For the merging system we have selected the innermost galaxies located inside the two circular regions ($R=18$ arcmin) centred on each BCG. The 132 spectroscopic cluster members have $\bar{z}=0.0461\pm0.0003$ and $\sigma_v/(1+\bar{z})=835\pm45$ \kms being, with 99\% c.l., Gaussian distributed (p-value~=~0.49 according to the Anderson-Darling test). The $\Delta$-test \citep{ds} points to the absence of  substructures with 99\% c.l. \citep[p-value~=~0.53;][]{hou12}.

We used the $n$-dimensional Gaussian mixture model algorithm implemented in the {\sc R} package {\sc Mclust} \citep{mclust}. The attempt to use the {\sc 1D-Mclust} to classify the galaxies according to their respective subcluster was, however, unsuccessful. According to the Bayesian Information Criterion (BIC), the preferred model (a single group) is strongly favoured over the second best model (two groups) with $\Delta{\rm BIC}=8$ \citep[][]{kass95}. The absence of large  substructures in the \red space is a hint that the merger axis is near to the  plane of the sky and/or the system is close to the most distant point in their orbit \citep[e.g.][]{monteiro-oliveira17,golovich16}.

\begin{figure*}
\begin{center}
\includegraphics[angle=-90, width=0.45\textwidth]{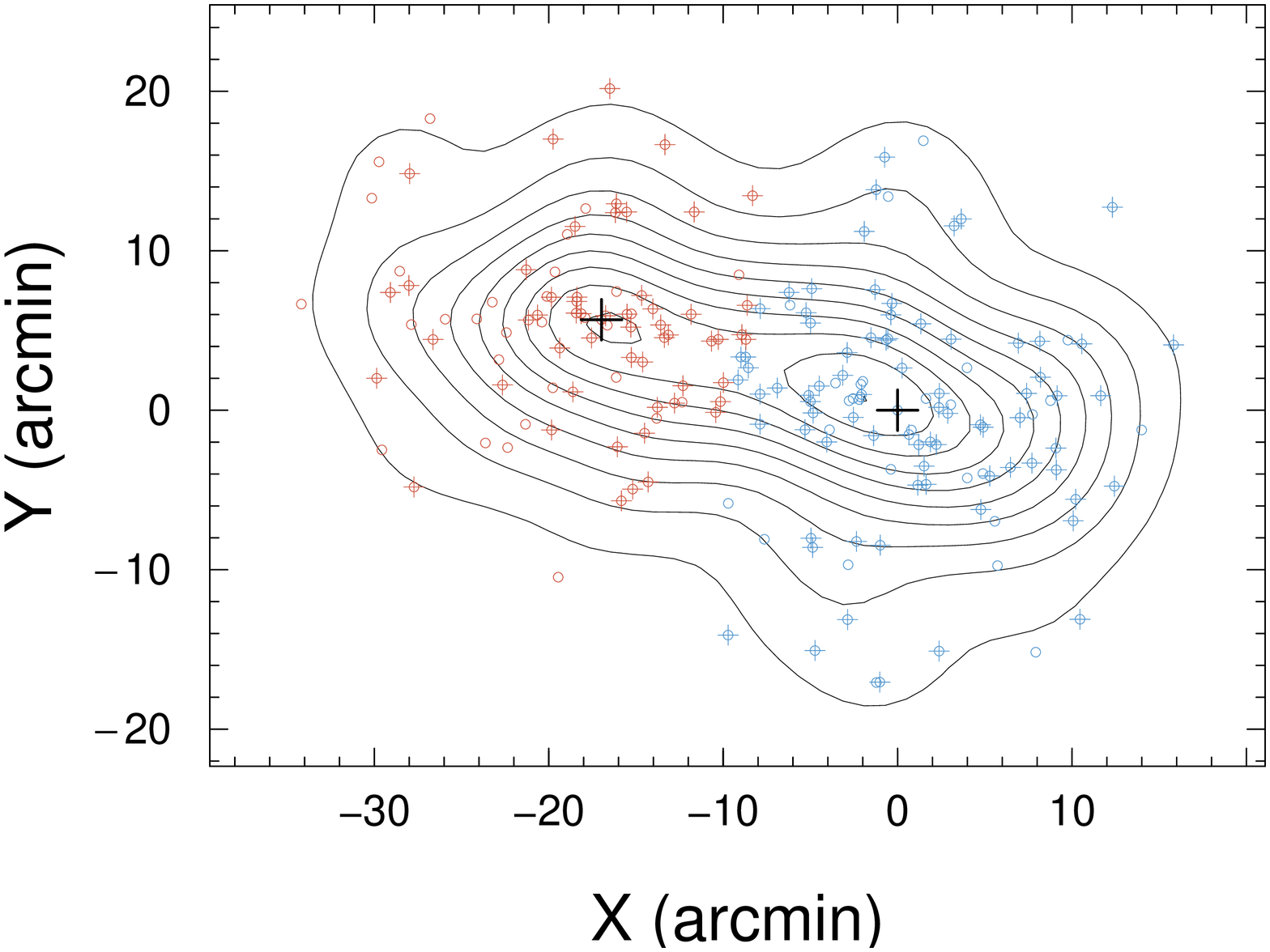}\quad 
\includegraphics[angle=-90, width=0.45\textwidth]{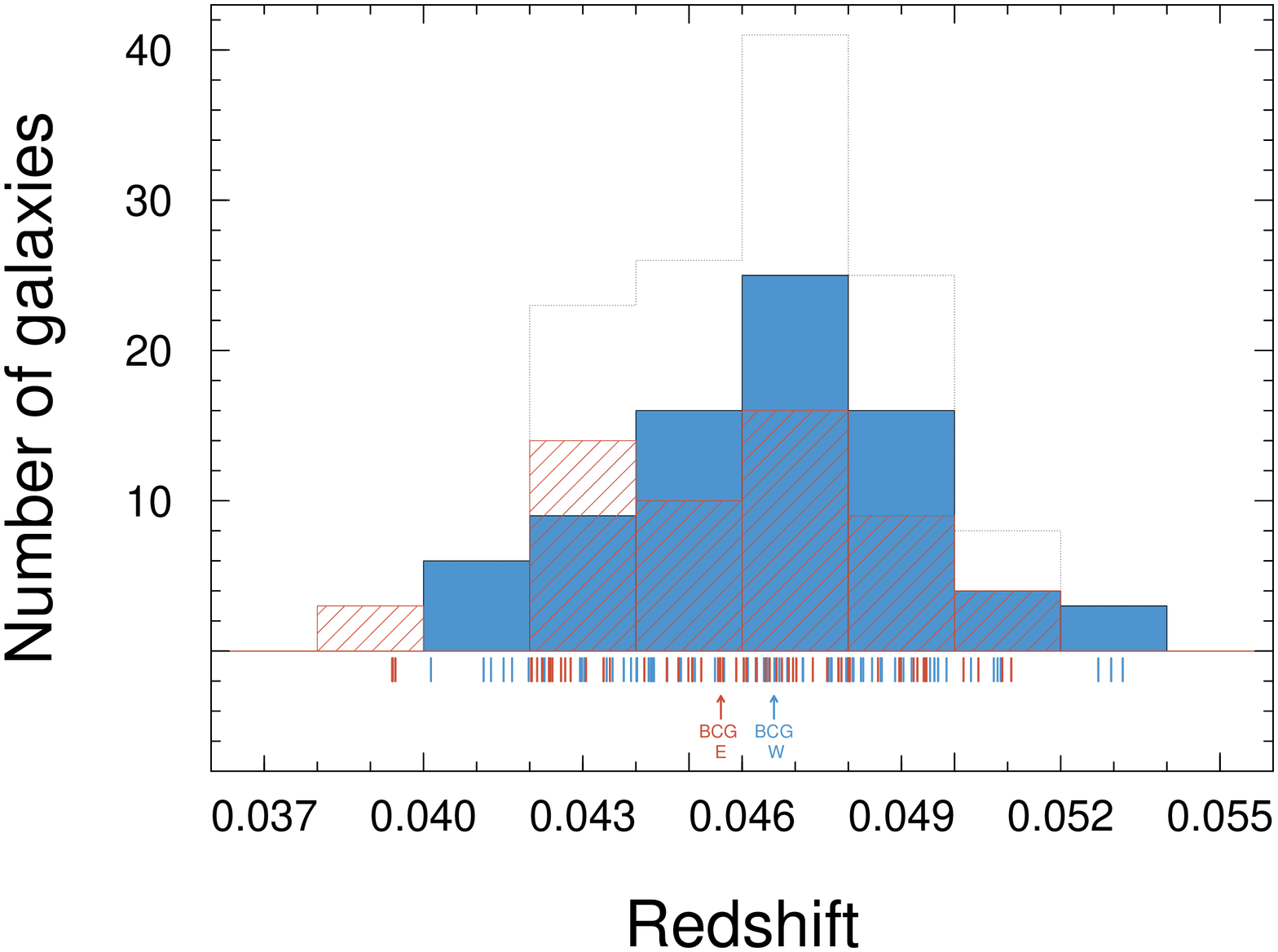}  
\caption{{\sc 2D-Mclust} results. {\it Left: } Spatial distribution of the subclusters A3376W ({\it blue}) \& A3376E ({\it red}) overlaid with the linear density contours for the 192 selected galaxies (continuous {\it black} lines). The ``+'' denotes the galaxies with available redshift, the {\it black} ones being the BCGs. {\it Right: } Total redshift distribution ({\it black} lines) overlaid with the recovered subclusters distribution following the same previous colour scheme. According to the Anderson-Darling test, both groups follow Gaussian distributions with 99\% c.l. Their \los separation is only $\delta v/(1+\bar{z})=181\pm147$ km s$^{-1}$.}

\label{fig:A3376:2d}
\end{center}
\end{figure*}

In order to overcome the impossibility of using the \reds as subcluster proxies, we have turned to the photometric red sequence galaxies. We have added the spectroscopic cluster members which are not initially in the sample for a total of 192 galaxies (70\% spectroscopic members). The {\sc 2D-Mclust} shows bimodality as the best result, with $\Delta{\rm BIC}=5$ in relation to the second best unimodal distribution. The recovered subclusters, shown in Fig.~\ref{fig:A3376:2d}, can be immediately related to the merging subclusters. Both are Gaussian with 99\% c.l. and are separated only by $\delta v/(1+\bar{z})=181\pm147$ \kms (68\% c.l.) along the \los. More details are presented in Tab.~\ref{tab:2dmclust}.

\begin{table}
\begin{center}
\caption{Dynamics of the subclusters A3376W \& A3376E modelled by {\sc 2D-Mclust}.}
\begin{tabular}{lcc}
\hline
\hline
	& A3376W & A3376E \\
\hline
Number of galaxies  	   & $79$	     & $56$ \\
$\bar{z}$  & $0.0464\pm0.0003$ & $0.0458\pm0.0004$ \\
$\sigma_v/ (1+\bar{z})$ (\kms) &  $815\pm60$   &$858\pm66$ \\
\hline
\hline
\end{tabular}
\label{tab:2dmclust}
\end{center}
\end{table}

The results are robust among several covariance matrix parametrisations implemented on {\sc Mclust}.  We also checked a possible contamination by galaxies belonging to the neighbour subcluster. We re-made the analysis only for the very innermost galaxies ($R\leq 8$ arcmin), and found similar results for the subclusters' mean \red as well $\delta v/(1+\bar{z})$.

The attempt to use the combined \red plus spatial coordinates ({\sc 3D-Mclust}) was also fruitless to disentangle the subclusters' galaxies ($\Delta{\rm BIC}~=~2.4$  relative to the second best bimodal solution). However, by setting $k=2$ groups, the algorithm brings the same result as its 2-dimensional version.

\section{Merger dynamics}
\label{sec:merger.scenario}

\subsection{A3376W \& A3376E}
\label{sec:merger.scenario.WE}

Our previous mass reconstruction and dynamical analysis enabled us to characterise the merger only at the observed time. In order to recover the A3376 merger history,  we have applied the  {\sc Monte Carlo Merger Analysis Code} \citep[MCMAC;][]{dawson}. This model presents an improvement with respect to the famous time argument model \citep{beers82}, because it provides the covariance estimation of the involved parameters (e.g. velocity, age and separation). The MCMAC considers a binary collision between two truncated NFW profiles. The input parameters, listed on Tab.~\ref{cap:A3376:tab:input.dawson}, were resampled through $2.5\times10^5$ realisations and applied to the model afterwards.

\begin{table}
\begin{center}
\caption{MCMAC input parameters based on our previous analysis. From the \reds we can estimate the current relative radial velocity of the subclusters, $v_{\rm rad}(t_{\rm obs})$ \citep[for more details see][]{dawson}. The projected  distance  $d_{\rm proj}$ corresponds to the mass peak separation.}
\begin{tabular}{lccc}
\hline
\hline
Parameter & Value & Uncertainty & Unit \\
\hline
M$_{200}^W$    & $3.0$     & $1.7$      & $10^{14}$ M$_{\odot}$ \\[2pt] 
M$_{200}^E$    & $0.9$     & $0.8$      & $10^{14}$ M$_{\odot}$ \\[2pt] 
$z_W$	       & $0.0464$  & $0.0003$    & --	\\[2pt] 
$z_E$	       & $0.0458$  & $0.0004$    & --\\[2pt] 
$d_{\rm proj}$ & $1096$    & $66$        & kpc\\[2pt] 
\hline
\hline
\end{tabular}
\label{cap:A3376:tab:input.dawson}
\end{center}
\end{table}

Normally,  MCMAC considers the merger direction equally probable between $0^\circ$ and $90^\circ$ from the plane of the sky. However, we have sufficient elements to better  constrain this value. The two subclusters are separated in \los by $181\pm147$ km s$^{-1}$. Concomitantly, the perpendicular velocity can be evaluated from the estimated shock velocity \citep[$v_{\rm shock}<2000$ km s$^{-1}$;][]{akamatsu12}. \cite{springel07} and more recently \cite{Machado+2015} have shown that halo velocity corresponds only to a fraction of the shock propagation. Considering $1000\pm500$ km s$^{-1}$ as the halo propagation velocity, we found a merger axis located at $\alpha=11^\circ \pm10^\circ$   with respect to the plane of the sky. Conservatively, we have adopted a prior in MCMAC, allowing $\alpha$ to assume any value between  $0^\circ$ and $20^\circ$ with equal probability.

The MCMAC results are shown in Tab.~\ref{cap:A3376:tab:output.dawson}. For comparison, the prior applied on $\alpha$ reduced by $\sim$58\% the error bar on the merger age ($TSC_0$) in relation to the unconstrained input.

\begin{table}
\begin{center}
\caption{MCMAC parameter estimation related to A3376W \& A3376E merger. The first three lines show the input quantities: $\alpha$ is the merger direction in relation to the plane of the sky, $d_{3D}(t_{\rm obs})=d_{\rm proj}/\cos \alpha$  is the current 3D distance between the subclusters and  $v_{3D}(t_{\rm obs})=v_{\rm rad}(t_{\rm obs})/\sin \alpha$  is the  current 3D relative velocity. The following lines present the quantities calculated by the model:  $d_{3D_{\rm  max}}$  is the 3-D maximum subcluster separation; $v_{3D}(t_{\rm col})$  is the 3D velocity at collision time; $v_{3D_{\rm max}}$  is the maximum relative velocity; $TSC_0$  is the time since collision for the outgoing scenario;  $TSC_1$  is the time since collision for the incoming scenario and $T$  is the collision period.}
\begin{tabular}{lccccccc}
\hline
\hline
Parameter &  Unit& Median & 68 \% c.l.  \\
\hline
$\alpha$	              & degree       &   11   &   8 -- 19 \\
$d_{3D}(t_{\rm obs})$     & Mpc          & 1.1    &   1.0 -- 1.2   \\
$v_{3D}(t_{\rm obs})$     & km s$^{-1}$  & 717    &   0 -- 991\\
$d_{3D_{\rm max}}$        & Mpc          & 1.4    &   1.0 -- 1.9    \\
$v_{3D}(t_{\rm col})$     & km s$^{-1}$  & 1878   &   1581 -- 2121 \\
$v_{3D_{\rm max}}$        & km s$^{-1}$  & 2477   &   2175 -- 2829 \\ 
$TSC_0$                   & Gyr          & 0.9    &   0.6 -- 1.1 \\
$TSC_1$                   & Gyr          & 2.5    &   1.1 -- 3.2  \\
$T$                       & Gyr          & 3.5    &   2.2 -- 4.1  \\
\hline
\hline
\end{tabular}
\label{cap:A3376:tab:output.dawson}
\end{center}
\end{table}

In the outgoing scenario, A3376W\&E are currently seen $0.9_{-0.3}^{+0.2}$ Gyr after the pericentric passage. This value is comparable, at 99\% c.l., with those previously estimated by \cite{akamatsu12}, \cite{machado13} and \cite{george15}. The collision velocity was $1878_{-297}^{+243}$ km s$^{-1}$, having been reduced to $717_{-717}^{+274}$ km s$^{-1}$ at the observation time. According to their relative position, the subclusters are nearly close to the apoapsis, $1.4\pm0.4$ Mpc, given their current separation of $1.1\pm0.1$ Mpc.

As an alternative scenario, we considered one in which the subclusters are returning after reaching maximum separation. The system is then seen $2.5_{-1.4}^{+0.7}$ Gyr after the collision. In the MCMAC sampling, $TSC_1$ is less than the age of the universe at the cluster redshift in $\sim$90\% of the realisations, in the sense that using only these results, we cannot discard the incoming scenario.

\begin{figure*}
 \begin{center}
 \includegraphics[angle=-90, width=0.45\textwidth]{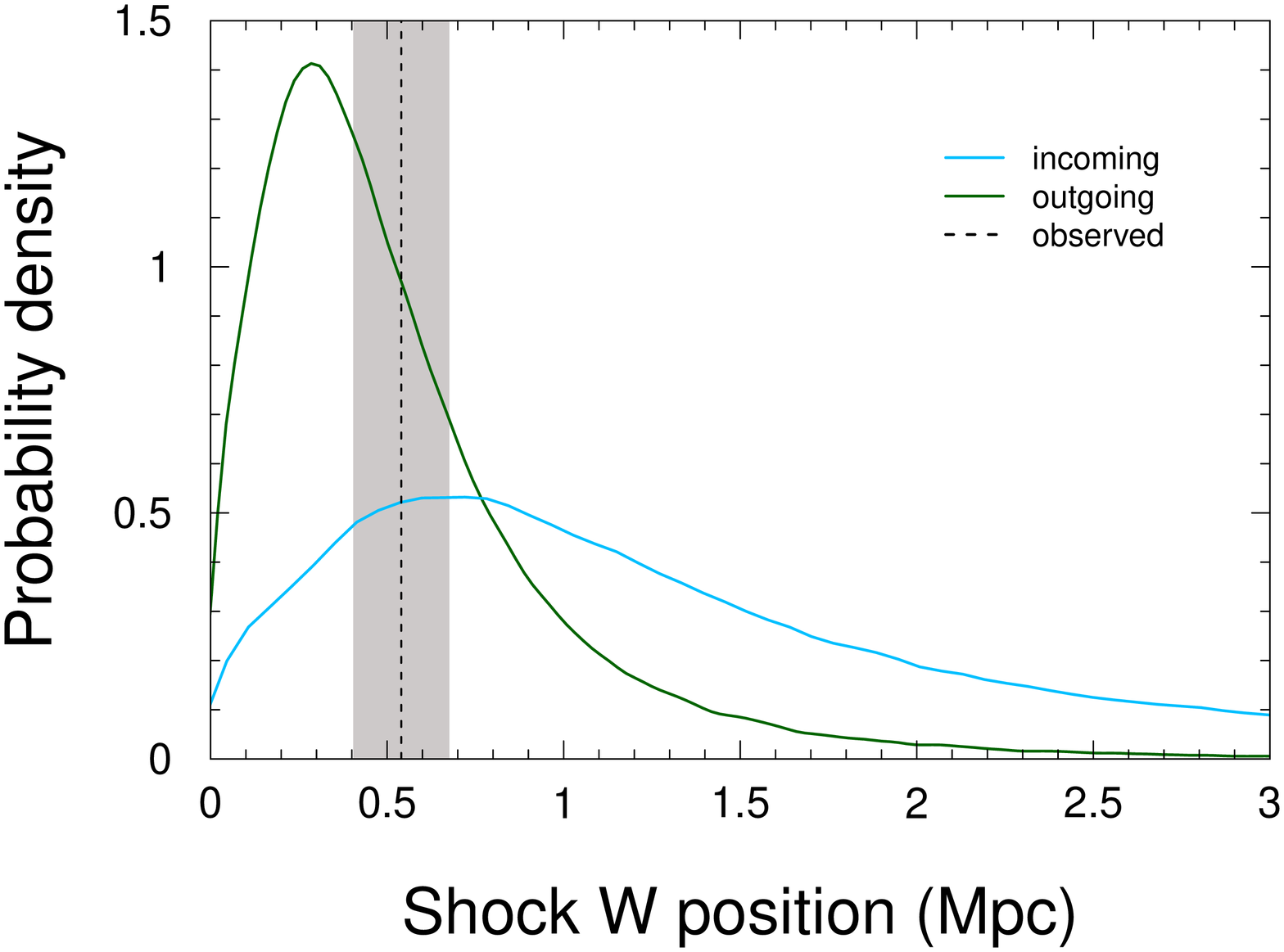}\quad 
\includegraphics[angle=-90, width=0.45\textwidth]{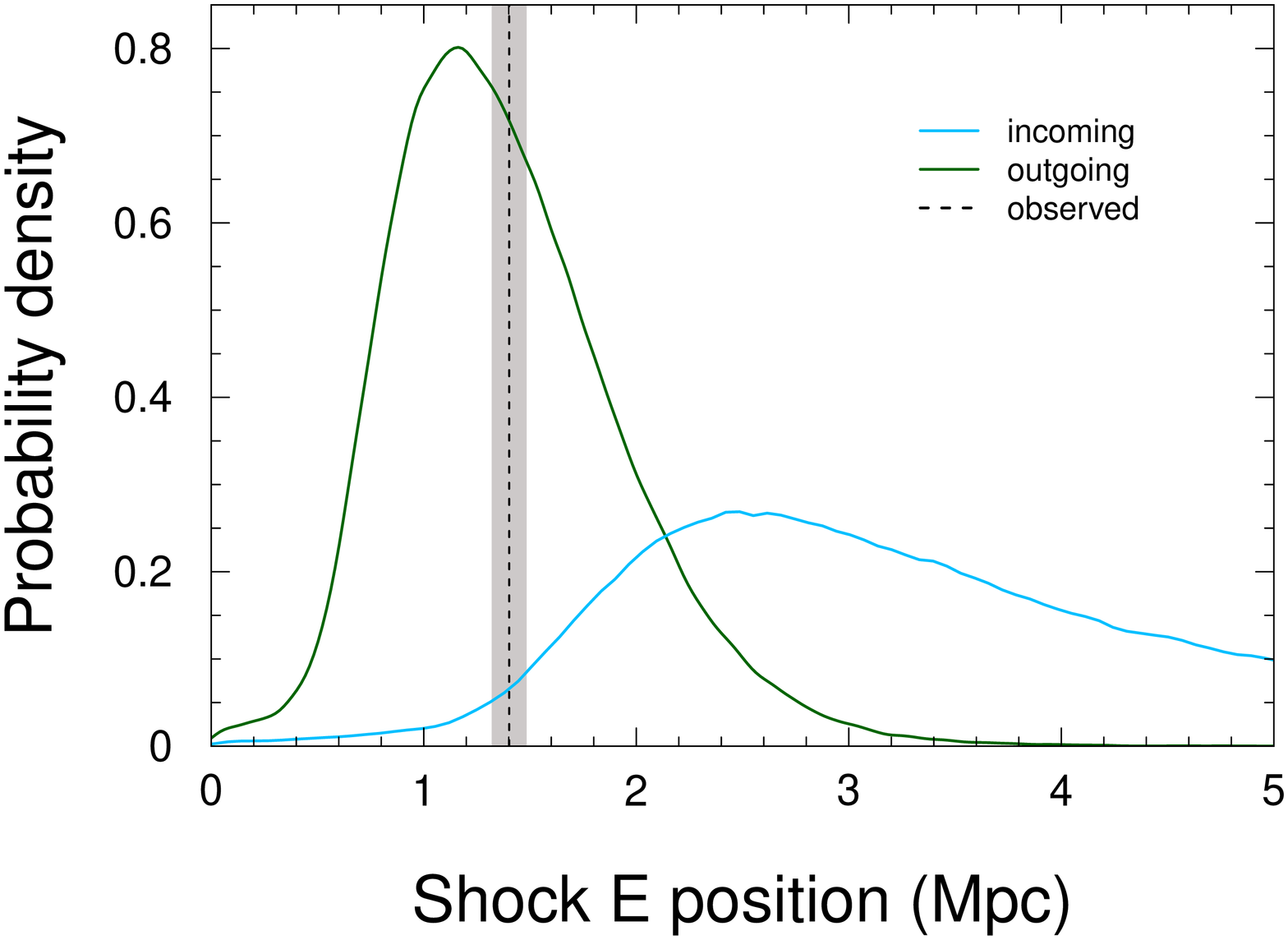}
 \caption{Predicted radio relics positions  with respect to the system mass centre calculated from the outputs of the two-body model. The outgoing and incoming posteriors are, respectively, the {\it green} and {\it cyan} lines. The measured relic position correspond to the {\it dashed} line where we highlighted in {\it grey} the interval of 68\% c.l., considered as the half of each relic width. While both outgoing and incoming scenarios are consistent within 68\% c.l. in the Western relics, the analysis of the  Eastern radio relic has showed that the outgoing scenario is $\sim$12 times more probable than the other one.}
 \label{fig:A3376:shock}
 \end{center}
 \end{figure*}

The model degeneracy can be broken by comparing each scenario prediction with an observational feature. One of the remarkable features in A3376 are the radio relics found in the outskirts of the system, which are thought to trace current shock positions. In relation to the mass centre of the system, the current shock position can be written as \citep{ng15}, 
\begin{equation}
s_i=\dfrac{M_j}{M_i+M_j}\ TSC \ v_{\rm shock}  \mbox{,}
\label{cap:A3376:eq:radio}
\end{equation}
where $i,j$ refer to the two subclusters, $M$ is the mass, $TSC$ is either outgoing or incoming age and $v_{\rm shock}$ is the shock velocity propagation. For this, we have chosen a uniform distribution $\mathcal{U}$(1600--2600 km s$^{-1}$) taking into account both observational \citep{akamatsu12} and numerical \citep{machado13}  estimates. The resulting expected shock positions are shown in  Fig.~\ref{fig:A3376:shock}.

Both relic positions argue in favour of the outgoing scenario. If the conclusion is not readily obvious in the W shock, where the models are compatible with 68\% confidence level, it is made explicit in the east shock: the outgoing scenario is $\sim$12 times more probable than the alternative one.

\subsection{A3376E \& A3376N}
\label{sec:merger.scenario.NE}

Although there is no observation of the ICM in the subcluster A3376N, we can conjecture about the interaction between it and the colliding system. A complete approach requires a complete three-body model. This, however, lies beyond the scope of this work. 

To study a possible interaction between A3376N \& A3376E separated by $1147\pm62$ kpc, a simple two-body only toy model can be adopted. We applied the generalised MCMAC to this effect,  which considers a possible unbound scenario \citep{andrade-santos15}. Also, in the absence of any merger angle prior, we have used the default uniform distribution\footnote{A possible merger between A3376E \& A3376N does not necessarily follow the same merger direction of the system A3376W \& A3376E}.

Our results show that in 34.4\% of $2.5\times10^5$ realisations, the model favours an unbound scenario. In this case, there is preference for small merger angles, $\alpha=5^\circ \pm4^\circ$.  Conversely, the bound solutions do not provide a good criterion to disentangle the outgoing or incoming scenario. In 34\% of the bound resulting realisations, $TSC_1=4.8^{+2.0}_{-3.1}$ Gyr is greater than the Universe age at the cluster redshift. For the outgoing configuration, the collision happened $1.5\pm1.0$ Gyr ago with $1849_{+234}^{-226}$ km s$^{-1}$. In the apoapsis the subclusters will then be separated by $2.7_{-1.6}^{+1.4}$ Mpc.

\section{Discussion}
\label{sec:discussion}

\subsection{A3376N}
\label{sec:disc.A3376N}

Both luminosity weighted density distribution and projected density maps point towards the presence of an additional structure in A3376 field, as suggested by \cite{ramella07}. Our dynamical analysis confirmed their compatibility with the main A3376 cluster. It is located at $\bar{z}=0.0472\pm0.0005$ having $\sigma/(1+\bar{z})=414\pm82$ km s$^{-1}$ and we report a mass of $1.4_{+0.7}^{-1.0}\times10^{14}$ M$_\odot$. We, however, must still be cautious, as  A3376N is under-sampled since these conclusions are based on just seven galaxies.

The absence of X-rays observation of its ICM hinder further detailed analysis of possible interactions between A3376N and A3376W\&E.

\subsection{Merger between A3376W \& A3376E}
\label{sec:disc.A3376WE}

\begin{figure}
 \begin{center}
 \includegraphics[angle=0, width=0.45\textwidth]{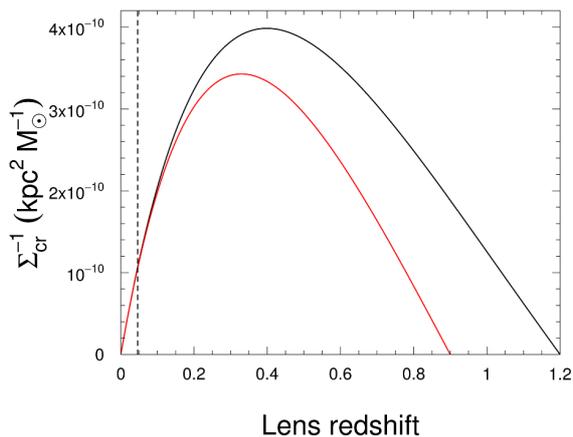} 
 \caption{Inverse of the critical surface density $\Sigma_{\rm cr}^{-1}$ as a function of the lens \red. The {\it black} line represents the case in which the background sources are fixed in $z_{\rm back}=1.2$ whereas in the {\it red} line the same galaxies are the closest, in $z_{\rm back}=0.9$. While the weak lensing signal ($\propto \Sigma_{\rm cr}^{-1}$)  remains nearly the same for a lens located at the A3376 redshift ($\bar{z}=0.046$, {\it dotted} line), the cut of higher \red sources acts to reduce the signal of higher \red lenses. In the absence of photometric redshifts, we tried to achieve the goal of removing the higher \reds lenses by cutting the fainter sources, most of them related to higher redshift.}
 \label{fig:A3376:lens.strength}
 \end{center}
 \end{figure}

The A3376 \red makes its weak lensing signal inescapably low. Therefore, background structures will appear that will have comparable or even higher S/N in the plane of the sky. Aiming to minimise this effect, we manipulated the cut off for background sample at the faint end, eliminating the most distant galaxies. In Fig.~\ref{fig:A3376:lens.strength} we can see an illustrative example of how the lensing signal changes when we move the background source cut from $z=1.2$ to $z=0.9$: while the signal of a lens at $z=0.046$ does not change, it is diminished for more distant lenses ($z>0.2$).

It is known that after galaxy cluster mergers, the BCGs will not be at rest with respect to the potential well \citep[generally coinciding with the centre of mass;][]{guo15}. This may, at least for some time, violate the central galaxy paradigm \citep[CGP;][]{bosch05} which could explain the apparent separation between the Northern BCG and its respective mass peak. However, \cite{guo15} have shown that this deviation, in general, is  more easily detected in velocity space than in real space. This corroborates our criteria to identify correlated mass concentrations as those closest to the BCGs.

The total mass of the system is $4.1^{+1.5}_{-1.8}\times10^{14}$ M$_{\odot}$. This value is, at 68\% c.l., in agreement with those estimated by \cite{girardi98}, based on the galaxy velocity dispersion. The subcluster's masses are M$_{200}^W=3.0_{-1.7}^{+1.3}\times10^{14}$ M$_\odot$ and M$_{200}^E=0.9_{-0.8}^{+0.5}\times10^{14}$ M$_\odot$, with  probability of 91\% for A3376W to be the more massive one, as predicted by \cite{machado13}.

The X-ray morphology reveals only one peak. The location of this X-ray peak coincides with the locations of both the Eastern BCG and of the corresponding mass peak with 95\% c.l. On the other hand, A3376W seems to have had its gas mostly stripped off. We have a configuration where one the two mass peaks does not coincide with a peak of X-ray emission. This makes A3376 a dissociative system. It is noteworthy that the most massive structure had its gas content dissipated, whereas the least massive one did not. One possible explanation is the scenario put forth by \cite{machado13}, in which the gaseous content of the Eastern subcluster is initially four times more concentrated than in the Western one. The smaller and denser A3376E passes through the more massive subcluster, disrupting its core. Although A3376W has its gas stripped, A3376E is able to preserve its denser gaseous core, as it emerges from the collision. That simulated scenario had arisen in an attempt to satisfy the observational constraints available then. Now, with the results of the present weak lensing analysis, additional evidence is presented in its favour. The current results confirm that the most important mass peak is indeed found towards the tail of the cometary shape, and not near the location of the X-ray peak.

Our dynamical analysis found that the subclusters are very close in the direction of the line-of-sight. The separation is only $\delta_v /(1+\bar{z}) = 181\pm 147$ km s$^{-1}$ which, combined with the estimated perpendicular velocity, yields a small merger angle of $10^\circ \pm11^\circ$. According to the two-body dynamical model, A3376W\&E are seen $\sim$0.8 Gyr after the pericentric passage. This value is consistent with previous estimations based on the shock position \citep{akamatsu12}, hydrodynamical simulations \citep{machado13} and the study of the radio relics \citep{george15}.  Despite the mass error bars being large, tests have shown that if we reduced them by 90\%, this would not bring significant differences on the estimated age. The more restrictive parameter estimation would come from the application of priors based on other observables \citep[e.g. X-ray temperature boost;][]{dawson}.

From the measured masses, we can verify the merger dynamical effect on the galaxy velocity dispersion by comparing their pre-merger values with those measured in the dynamical analysis \citep{monteiro-oliveira17}. From the scaling relations proposed by \cite{biviano06}, we have found a pre-merger value of $750_{-128}^{+134}$ km s$^{-1}$ and $505_{-121}^{+125}$ km s$^{-1}$, for A3376W and A3376E respectively, under mass conservation during the collision. Defining the velocity dispersion boost factor as  $f\equiv\sigma_{\rm obs}/\sigma_{\rm pre}$, we found  $f_{\rm W}=1.1^{+0.2}_{-0.3}$ and  $f_{\rm E}=1.7^{+0.4}_{-0.5}$.  Since these two values are greater than one, a post-collisional scenario is favoured in both cases \citep[see Fig.~29 of ][]{pinkney}.

Our general results corroborate the merger scenario previously proposed by \cite{machado13}. The weak lensing analysis suggests two mass concentrations around the two BCGs. Based on the two-body dynamical model, we have found that A3376W\&E is seen after $0.9_{-0.3}^{+0.2}$ Gyr after the pericentric passage and are currently going to the orbit apoapsis.
On the other side, the present work suggest that the merger direction is only $11^\circ \pm10^\circ$ from the plane of the sky, whereas the previous simulations prefer a value of $40^\circ$. 

\section{Summary}
\label{sec:summary}

\begin{itemize}

    \item Our large field-of-view images have enabled us to confirm that A3376 is, actually, a trimodal system. The ``new'' structure, located $\sim$1150 kpc north-east from A3376E, has a mass $1.4_{-1.0}^{+0.7}\times10^{14}$ M$_{\odot}$. The seven spectroscopic member galaxies show that the structure has a mean redshift of $0.0472\pm0.0005$ and a velocity dispersion of $414\pm82$ km s$^{-1}$;
    
    \item Despite the challenges to the task, posed by the very low redshift of A3376, we have succeeded in determining the merging subclusters' mass distributions. We found the individual masses  M$_{200}^W=3.0_{-1.7}^{+1.3}\times10^{14}$ M$_\odot$ and M$_{200}^E=0.9_{-0.8}^{+0.5}\times10^{14}$ M$_\odot$ leading to a ratio  $M_W/M_E=3.3_{-3.0}^{+2.0}$. Additionally, we found that there is a probability of 91\% that A3376W is the most massive subcluster;
    
    \item The X-ray morphology presents only one X-ray peak, related to A3376E. It is, with 95\% c.l., in agreement with both BCG and mass peak positions. On the other hand, the gas content of A3376W seems to have been stripped out. 
    
    \item Our cluster member classification based on galaxy spatial distribution plus redshift points out that A3376W\&E are located close to each other in relation to the \los. The velocity difference is only $\delta_v /(1+\bar{z}) = 181\pm 147$ km s$^{-1}$; combined with an estimation of the perpendicular velocity based on the shock propagation, we found that the merger is taking place just at $11^\circ \pm10^\circ$  from the plane of the sky;
    
    \item According to the two-body model, the collision occurred $0.9_{-0.3}^{+0.2}$ Gyr ago with a velocity of $1849_{-226}^{+234}$ km s$^{-1}$. Notwithstanding the model degeneracy, the radio relic positions support the outgoing scenario.
\end{itemize}

\section*{Acknowledgements}
\addcontentsline{toc}{section}{Acknowledgements}

The authors thank the referee for providing helpful suggestions that improved the paper. RMO thanks Marcelle Soares-Santos and Alistair Walker for valuable advices and Andr\'e Zamorano Vitorelli for the language revision. RMO also thanks the finantial support provided by CAPES and CNPq (142219/2013-4) and the project {\it Casadinho} PROCAD - CNPq/CAPES (552236/2011-0). GBLN acknowledges partial support form CNPq. ESC and REGM acknowledges support from FAPESP (grants: 2014/13723-3 and 2010/12277-9, respectively). TFL acknowledge financial support from FAPESP and CNPq (grants: 2012/00578-0 and 303278/2015-3, respectively).

This work is based in part on data products produced at Terapix available at the Canadian Astronomy Data Centre as part of the Canada-France-Hawaii Telescope Legacy Survey, a collaborative project of NRC and CNRS. 

This work has made use of the computing facilities of the Laboratory of Astroinformatics (IAG/USP, NAT/Unicsul), whose purchase was made possible by the Brazilian agency FAPESP (grant 2009/54006-4) and the INCT-A. 

We made use of the NASA/IPAC Extragalactic Database, which is operated by the Jet Propulsion Laboratory, California Institute of Technology, under contract with NASA.








\bibliographystyle{mnras}
\bibliography{bibliografia}



\bsp	
\label{lastpage}
\end{document}